\documentclass[conference,compsoc]{IEEEtran}

\ifCLASSOPTIONcompsoc
  \usepackage[nocompress]{cite}
\else
  \usepackage{cite}
\fi

\usepackage{amsmath,amssymb,amsfonts}
\usepackage{graphicx}
\usepackage{xcolor}
\usepackage{colortbl}
\usepackage{booktabs}
\usepackage{array}
\usepackage{tabularx}
\usepackage{placeins}
\usepackage[hyphens]{url}
\let\codeurl\url
\usepackage{microtype}
\usepackage{fontawesome}

\definecolor{LMSMBlue}{HTML}{34D0F4}
\definecolor{LMSMGreen}{HTML}{FA7299}
\definecolor{LMSMGreenDark}{HTML}{B0175A}
\definecolor{LMSMCheckpoint}{HTML}{B0175A}
\definecolor{LMSMAnytime}{HTML}{FA7299}
\definecolor{LMSMDisabled}{HTML}{BFC4D6}
\definecolor{LMSMText}{HTML}{16204A}

\definecolor{LMSMBlueTint}{HTML}{EAF7FD}
\definecolor{LMSMLink}{HTML}{0F6B8A}
\definecolor{LMSMGreenTint}{HTML}{FCEBF2}

\newcommand{\tblswatch}[1]{%
  \raisebox{0.15ex}{\textcolor{#1}{\rule{0.85ex}{0.85ex}}}%
}

\graphicspath{{figures/}}
\newcommand{\lmsmnativedots}{blue}

\usepackage[colorlinks=true,linkcolor=LMSMLink,citecolor=black,urlcolor=LMSMLink,bookmarksnumbered=true,
  pdftitle={Language Model Security Modules: Turning LLM Interpretability into Runtime Defense},
  pdfauthor={XiuYu Zhang, Bonan Ruan, Junfeng Fang, An Zhang, Tat-Seng Chua, Zhenkai Liang}]{hyperref}
\begin{document}
\bstctlcite{IEEEexample:BSTcontrol}
\title{Language Model Security Modules:\\ Turning LLM Interpretability into Runtime Defense}

\author{\IEEEauthorblockN{XiuYu Zhang\textsuperscript{1},
Bonan Ruan\textsuperscript{1},
Junfeng Fang\textsuperscript{1},
An Zhang\textsuperscript{2},
Tat-Seng Chua\textsuperscript{1}, and
Zhenkai Liang\textsuperscript{1}}
\IEEEauthorblockA{\textsuperscript{1}National University of Singapore\qquad
\textsuperscript{2}University of Science and Technology of China\\[0.5em]
{\upshape\faGithub\ \href{https://github.com/xiuyuz/LMSM}{github.com/xiuyuz/LMSM}}}}

\maketitle
\thispagestyle{plain}

\begin{abstract}
Large language models (LLMs) are served behind layered defenses such as alignment and guardrail models, yet jailbreaks still get through.
Interpretability research looks inside the model, where methods such as probes and sparse autoencoders expose signals of harmful content during generation.
A signal alone, however, is not a defense.
To become one, the serving system must tie each signal to the right request, apply the policy as declared, and release output only when the policy allows.
Today, defenses built on such signals handle these duties on their own, so every new detector needs its own integration.
Yet none of these duties depends on how a signal is computed.
We present Language Model Security Modules (LMSM), which adapts the design of Linux Security Modules to LLM serving and carries out these duties once in a shared mediation path.
The path hosts pluggable backends that turn model internals into evidence and policies that decide what that evidence allows.
Each policy declares the evidence and evaluations it needs, and LMSM provides exactly these for each request and verifies that they took place, so a failure is never mistaken for permission.
Protection therefore remains cheap and correct under common serving optimizations.
We evaluate LMSM in vLLM, with both eager and compiled execution, on Qwen3-4B and Gemma-3-4B with fitted probes, a released transcoder, and a released sparse autoencoder.
With vLLM's default optimizations on, every backend keeps at least 96\% of native throughput, and all of them share the same mediation path.
On Qwen3-4B, a representative probe-based policy cuts HarmBench attack success from 39.4\% to 3.3\% while XSTest false refusals rise from 0.8\% to 3.1\%.
LMSM lets existing and future interpretability methods become runtime defenses without rebuilding the serving path.
\end{abstract}

\IEEEpeerreviewmaketitle

\section{Introduction}
\label{sec:introduction}

Large language models (LLMs) are no longer deployed as isolated predictors that answer one prompt and disappear.
They run inside serving systems that answer many users at once.
The organizations that operate these services, which we call operators, rely on several layers of defense to keep restricted content out of their responses.
Alignment shapes the model's default behavior~\cite{ouyang2022instructgpt,zhang2026alphaalign}, prompt-level controls place instructions in its context~\cite{wallace2024instruction}, and guard models inspect the text that goes in and out~\cite{inan2023llama,rebedea2023nemo}.
Malicious requests, however, still get through these layers~\cite{zou2023universal,mazeika2024harmbench,jiang2024wildteaming}.

Recent work has begun to explore another line of defense that operates inside the model while it generates.
Probes, sparse autoencoders (SAEs)~\cite{huben2024sparse}, transcoders~\cite{dunefsky2024transcoders}, and related interpretability methods read the activations that the model uses to produce its next token and expose signals that track harmful intent or content~\cite{arditi2024refusal,zhang2025jbshield,jiang2025hiddendetect}.
This line of defense benefits directly from progress in interpretability.
As researchers discover more stable internal features and more faithful detectors of harmful concepts, those discoveries can become runtime safety signals.
Providers have started to use such signals in production, where activation probes are far cheaper than separate guard models~\cite{cunningham2026constitutional,kramar2026probes}.

\begin{figure*}[t]
    \centering
    \includegraphics[width=0.95\textwidth]{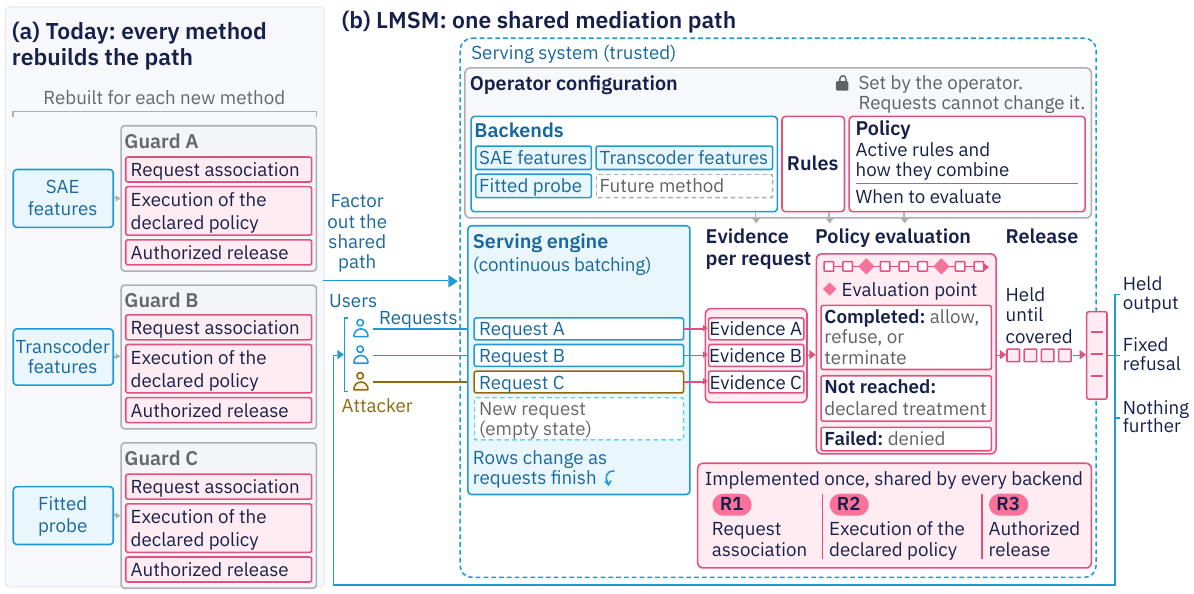}
    \caption{
        \textbf{LMSM turns model-internal evidence into an enforced policy through one shared mediation path.}
        (a) Today, each method that adapts model-internal evidence for safety handles request association, execution of the declared policy, and authorized release on its own.
        (b) In LMSM, the operator installs backends, rules, and a policy that requests cannot change.
        The shared path keeps each request's evidence with that request (R1), performs the evaluations that the policy declares and records their outcome (R2), and releases output only after the evaluations that cover it have completed (R3).
    }
    \label{fig:lmsm-vision}
\end{figure*}

However, a useful signal is not yet a defense.
Consider an operator that must refuse responses giving working instructions for breaking into computer systems, a restriction that its aligned model does not reliably honor.
A transcoder feature or a fitted probe may respond reliably when the model begins to write such instructions.
To turn that response into protection, the serving system must still perform three duties that we describe as requirements R1 to R3.
It must attach the evidence to the right request while it serves many users in shared batches (R1), evaluate the operator's rules at the moments the policy requires (R2), and release a response only as those evaluations allow (R3).

Although none of these requirements depends on how the signal is produced, the solutions adapting model-internal evidence for safety handle these requirements in ad-hoc ways, each tightly coupled to one artifact, one calibration procedure, one harm category, and one intervention technique (Figure~\ref{fig:lmsm-vision}a).
These solutions show that useful internal signals exist, but they do not provide a general solution for turning those signals into deployed protection.
After a better interpretability solution is developed, the full solution needs to be completely redeveloped to adopt it, which makes maintenance a daunting task. 

Systems security once faced a similar problem when building security solutions in the Linux kernel, where the reference monitor principle requires every security-relevant operation to pass through a small trusted mediator~\cite{anderson1972reference,saltzer1975protection}.
Linux Security Modules (LSM) made this practical in the kernel by providing stable mediation points and letting pluggable modules supply the policy~\cite{wright2002lsm}, instead of adding ad-hoc functionalities in the Linux kernel.
The lasting value of LSM is not any single policy, such as SELinux~\cite{loscocco2001selinux} or AppArmor~\cite{apparmorAppArmor}, but a framework in which policies can change while the mediation path remains the same.
Recent work brings these principles to LLM agents~\cite{christodorescu2026agentsecurity,christodorescu2025sok,wu2024newera}, treating the model as untrusted and mediating what an agent does with its tools through isolation~\cite{wu2025isolategpt}, capabilities~\cite{debenedetti2025camel}, information-flow control~\cite{costa2025fides}, and least privilege~\cite{shi2025progent}.

Inspired by the security principles that LSM put into practice, we present Language Model Security Modules (LMSM), a framework that implements the path from evidence to release once and lets backends and policies change around it (Figure~\ref{fig:lmsm-vision}b).
The operator installs backends, which wrap interpretability methods and turn activations into evidence, and a policy, which decides from that evidence how each request is handled.
As in LSM, the name refers to the framework, and the modules it hosts are the backends and policies.
LMSM keeps each request's evidence attached to that request, runs the evaluations that the policy declares and records whether each one completed, and lets output reach the requester only as the outcome of those evaluations allows.
If a new SAE exposes a better concept, a transcoder captures a more useful computation, or a new probe offers a cleaner signal, the operator should need only a backend adapter and calibrated rules to deploy it.
Our central argument is that progress in interpretability should translate directly into stronger runtime defenses, and that this requires a mediation path built for that purpose.

This design also lets protection run at the speed of modern serving.
Because each policy declares the evidence it uses and when to evaluate it, LMSM computes only that evidence.
A policy built on a released transcoder, for example, reads 15 of its 163,840 features.
LMSM keeps all state with requests rather than batch positions and checks every declared evaluation.
As a result, a serving engine can keep optimizations such as continuous batching and compiled execution without risking a silent bypass.

We implement LMSM in Hugging Face Transformers~\cite{wolf2020transformers} and in vLLM~\cite{kwon2023vllm}.
Our evaluation asks separately whether LMSM applies each policy correctly and how well the policy stops harmful output.
In tests where requests constantly join and leave shared batches, we observed no evidence or decision reaching the wrong request, and each request whose evidence we removed was refused.
LMSM hosts a released SAE on Gemma-3-4B and a released transcoder on Qwen3-4B, and replacing fitted probes with the transcoder leaves all three requirements unchanged.
In compiled vLLM with its default optimizations on, each backend we tested keeps at least 96\% of native throughput, while the integrations that other designs require keep 17 to 85\%.
Averaged over three seeds, a representative policy reduces the attack success rate on HarmBench~\cite{mazeika2024harmbench} from 39.4\% to 3.3\%, while false refusals on XSTest~\cite{rottger2024xstest} rise from 0.8\% to 3.1\%.
On WildJailbreak~\cite{jiang2024wildteaming}, whose training split we used to fit the probes, it reduces the attack success rate from 43.2\% to 6.7\%.

In summary, we make three contributions.
\begin{itemize}
    \item \textbf{Formulation.} We identify the three duties required to turn model-internal evidence into an enforced operator policy.
    They are request association, execution of the declared policy, and authorized release, and we show that none of them depends on how the evidence is produced.

    \item \textbf{Architecture.} We design LMSM, a reusable mediation path in which each policy declares the evidence and evaluations it requires and every request records whether they ran.
    Backends and policies can then change without rebuilding the shared mediation path, and serving engines keep their optimizations.

    \item \textbf{Validation.} We implement LMSM in Hugging Face Transformers and in compiled vLLM with continuous batching.
    We show that it meets all three requirements in the serving configurations it supports and that it hosts both released interpretability artifacts and probes fitted for the deployment.
    Without keying state by request and resetting it at admission, ordinary batching releases 878 HarmBench responses that the same policy otherwise refuses.
    Policies on each backend we tested keep at least 96\% of native throughput under vLLM's default optimizations, and the probe policies remove most harmful output at a small cost in false refusals.
\end{itemize}

\section[From Layered Defenses to Runtime Mediation]{From Layered Defenses\\ to Runtime Mediation}
\label{sec:motivation}

Where a defense acts between a request and its response decides what it can observe and what it can still prevent.
We first compare the layers of defense that operators combine and show what the model-internal layer adds.
We then explain why this layer still lacks a shared mediation path and what LSM teaches about building one.

\subsection{Layers of Defense}
\label{sec:bg-layers}

Operators combine several layers of defense because each layer acts at a different point and sees different information (Table~\ref{tab:defense-landscape}).
External guards check the text that enters and leaves the model with a classifier of their own.
They are independent of the model and easy to update, but an output guard can judge a response only after the model has written it.

\begin{table}[t]
  \centering
  \caption{
    Layers of defense around an LLM.
  }
  \label{tab:defense-landscape}
  \footnotesize
  \setlength{\tabcolsep}{3pt}
  \renewcommand{\arraystretch}{1.10}
  \begin{tabularx}{\columnwidth}{
    >{\raggedright\arraybackslash}p{0.31\columnwidth}
    >{\raggedright\arraybackslash}X
    >{\raggedright\arraybackslash}X
  }
    \toprule
    \textbf{Layer and where it acts} & \textbf{Strength} & \textbf{Limitation} \\
    \midrule
    \tblswatch{LMSMBlue}\ \textbf{External guards}~\cite{inan2023llama,rebedea2023nemo,chennabasappa2025llamafirewall}
      \newline \emph{Text entering and leaving the model}
      & Independent of the model and easy to update
      & Sees only text, and a response only after the model has written it \\
    \midrule
    \tblswatch{LMSMBlue}\ \textbf{Prompt controls}~\cite{wallace2024instruction}
      \newline \emph{Model context, before generation}
      & No change to the model or the serving system
      & Instructions share the context with the attacker's text \\
    \midrule
    \tblswatch{LMSMBlue}\ \textbf{Alignment}~\cite{ouyang2022instructgpt,rafailov2023direct,bai2022constitutional}
      \newline \emph{Model weights, before deployment}
      & Broad default behavior for all requests
      & Set by the model provider, and an operator's own restriction needs further training \\
    \midrule
    \tblswatch{LMSMBlue}\ \textbf{Model-internal defenses}~\cite{zhang2025jbshield,jiang2025hiddendetect,lee2025cast}
      \newline \emph{Activations, during generation}
      & Sees the computation behind each token before the token is released
      & Each method brings its own decision logic and action, and request association and release are left to the operator \\
    \midrule
    \rowcolor{LMSMGreenTint}
    \tblswatch{LMSMGreen}\ \textbf{LMSM}
      \newline \emph{Model-internal evidence, during generation}
      & One shared mediation path for any backend and policy
      & Requires a trusted serving system and calibrated rules for each backend \\
    \bottomrule
  \end{tabularx}
\end{table}

Prompt-level controls and alignment act before generation, on the model's context and weights, and they shape what the model tends to do rather than check what it does.

Model-internal defenses act during generation.
They read the activations from which the model computes each next token, and interpretability methods turn these activations into evidence.
A probe is a small classifier fitted to activations labeled with a behavior~\cite{alain2016probes}.
An SAE~\cite{huben2024sparse} or a transcoder~\cite{dunefsky2024transcoders} rewrites an activation as a combination of many sparse features, many of which respond to a single concept.
A rule can name such a feature and compare its value with a calibrated threshold.
Internal evidence describes the computation behind a token before the token is released, and it comes from the model's own forward pass rather than from a second model that reads the text.

Moving inward does not rank the layers from weak to strong.
Internal evidence exists only for behaviors that the model represents in its activations.
An external guard brings knowledge of its own, and alignment sets the default behavior for all requests.
Therefore, LMSM strengthens the model-internal layer and leaves the other layers in place.

\subsection{The Missing Mediation Path}
\label{sec:bg-gap}

Existing model-internal defenses show that activations reveal harmful requests~\cite{zhang2025jbshield,jiang2025hiddendetect,lee2025cast}.
Each of them defines when its evidence is evaluated and what follows, from marking a request as unsafe to steering activations toward a safe answer.
None of them describes how evidence remains with its request when a serving system batches many requests, or how output is held until the evaluation has run.
Reports of probes deployed in production do not describe these duties either~\cite{cunningham2026constitutional,kramar2026probes}, which leaves them to each operator.

Engine instrumentation works from the serving side.
The MIA plugin of vLLM Hook~\cite{fang2026mia}, DMI-Lib~\cite{yu2026dmilib}, and vLLM-Lens~\cite{cooney2026vllmlens} capture activations for each request in a batch.
What the captured evidence means is left to the analysis code that receives it, and whether output waits for that analysis is left to the operator.
For text, guardrail frameworks already host many checks behind one interface~\cite{rebedea2023nemo,chennabasappa2025llamafirewall}.
Internal evidence has methods that produce it and tools that capture it, but no shared mediation path that keeps it with each request, executes the declared policy over it, and releases output only as that policy allows.

\subsection{Lessons from LSM}
\label{sec:bg-lsm}

The Linux kernel faced a similar situation more than two decades ago.
Many projects had built stronger access control for Linux as separate kernel patches, and the security community did not agree on which model was right~\cite{wright2002lsm}.
Rather than choose among the competing projects, the kernel developers asked for an infrastructure that lets each model load as a module, and LSM was built to provide it.
LSM places its hooks in the kernel just before each access to an internal object, such as a file or a task, rather than at the system-call boundary.
Mediation at that boundary may duplicate kernel code and miss the context of the decision~\cite{wright2002lsm}.
External guards resemble mediation at that boundary.
A guard judges the text with a model of its own, whereas a backend reads the computation that produced the text.

At each hook, the kernel passes the context of the operation to the module, the module decides under the policy it was loaded with, and the kernel enforces the decision.
The kernel side ensures that each decision concerns the right process and object, that the module is consulted on each mediated operation, and that the operation proceeds only if the module allows it.
These are the three duties that the model-internal layer lacks, and the kernel handles them once for all modules.

LMSM applies this division to LLM serving.
A backend supplies evidence, a policy decides what the evidence allows, and the shared mediation path handles the three duties once for all backends and policies.
The analogy also has limits.
A serving system is not a kernel, and LMSM does not inherit kernel isolation or tamper resistance, which our threat model states as trust assumptions.
A kernel hook also runs because it lies on the code path that it guards.
In a serving engine, one forward pass computes the evidence of many requests, and engine optimizations can keep an observer from running.
Therefore, LMSM checks for each request that the declared evaluations took place.
Finally, a hook sees the operation it mediates, whereas a backend reports only a statistical signal about the computation.
As a result, how well a policy recognizes harmful output is a separate question, and our evaluation answers it.

\section{System and Threat Model}
\label{sec:security-model}

This section describes the deployments LMSM targets, the attacker it defends against, and the requirements a serving system must meet to turn model-internal evidence into an enforced policy.
Two questions run through the paper.
The first is whether the serving system applies the operator's policy correctly.
The second is whether the policy, given its evidence, recognizes the behavior the operator wants to stop.
LMSM answers the first question, and our evaluation measures the second for the policies we deploy.

\subsection{System Model}
\label{sec:deployment-model}

We consider a service that hosts an aligned language model and answers requests from many users.
The operator that runs the service may not be the model provider and often has restrictions of its own.
In our running example, it wants to refuse responses that give working instructions for breaking into computer systems, even when the model itself does not refuse.
To enforce such a restriction, the operator installs one or more \emph{backends} alongside the model.
A backend reads the model's activations during generation and reports evidence about whether a restricted behavior is present.
It may reuse a released interpretability artifact, such as an SAE or a transcoder, or use a probe that the operator trains itself.
The operator writes \emph{rules} over this evidence, and each rule pairs a calibrated condition with an action, such as refusing the request.
A \emph{policy} selects which rules are active and how their outcomes combine into one decision for a request.
It also states when the rules are evaluated during generation and declares how the service treats a request whose evaluation is never reached.

The service runs on a serving engine that uses continuous batching, the standard technique for high-throughput serving~\cite{kwon2023vllm}.
At each step, the engine advances all active requests in one forward pass and admits new requests as soon as finished ones free capacity.
As a result, evidence is computed for many requests at once, and the place a request occupies in the batch changes while it is being generated.

The operation LMSM protects is the \emph{release} of generated text to the requester.
Tokens, activations, and other intermediate state stay inside the serving system until they are released.
A service can release a response all at once or token by token.
If the policy refuses a request, the service releases a fixed refusal chosen by the operator instead of the held output.
If the policy terminates a request, the service releases no further output.

\subsection{Threat Model}
\label{sec:threat-model}

\textbf{Attacker.}
The attacker is a user of the service.
It writes the requests it submits and may use any prompting strategy, including known jailbreak techniques that try to elicit restricted content.
It may also send many requests at the same time and vary their lengths, which changes how the engine batches requests and when it reuses the capacity of finished ones.
The attacker wins if the service releases a response that the operator's policy prohibits, or if its requests cause the decision for another user to rely on evidence or state from a different request.
The attacker knows how LMSM works, since its guarantees do not rely on a secret design.

\textbf{Trust assumptions.}
We trust the operator to configure the service as intended, and we assume that the components it installs are not compromised.
These components are the model, the serving engine, the backends, the policy, and LMSM itself.
They run inside the serving system, much like inlined reference monitors~\cite{erlingsson2000irm}.
Requests can change what the model generates, but they cannot choose or change backends, rules, or policies.
In the deployments we consider, the operator also sets the generation parameters.
The serving engine assigns each active request a unique identifier, and LMSM relies on it to tell requests apart.
Attacks on the serving system itself and attacks that only reduce availability are outside our scope.

\textbf{Scope.}
We study single-turn requests.
Conversation history, retrieved documents, and tool outputs can also contain attacker-controlled content, and we leave them to future work.
LMSM decides how evidence is applied, not how hard the evidence is to evade.
Prior work shows that inputs crafted against a specific monitor can evade it~\cite{bailey2026obfuscated}.
Therefore, we measure the effectiveness of our policies on standard jailbreak benchmarks and treat robustness against attackers who adapt to a particular backend as a property of that backend.
Under default GPU kernels, the numbers that a forward pass computes for a request can depend slightly on which other requests share the batch.
This affects the tokens that the model generates as well as the evidence that backends compute, and a decision that lies very close to its threshold can change with it.
Such a decision still relies on the request's own evidence.
The change comes from the engine's arithmetic, and no evidence passes from one request to another.
vLLM offers a batch-invariant mode that removes this effect at a throughput cost.
LMSM runs in both modes wherever vLLM supports them, and we evaluate it in both.

\subsection{Mediation Requirements}
\label{sec:security-objectives}

We now state the three duties from the introduction as requirements R1 to R3 on the serving system, and Figure~\ref{fig:lmsm-vision}b shows where each one applies.

\textbf{Request association (R1).}
Evidence, accumulated policy state, and decisions must stay with the request that produced them.
Continuous batching makes this harder than it sounds.
A request's evidence arrives as one row of a shared batch whose layout changes as requests finish and new ones take their place, and a newly admitted request must start without any state left by the request it replaced.

\textbf{Execution of the declared policy (R2).}
The serving system must perform the observations and evaluations that the policy specifies and record for every request whether they took place.
Some requests cannot be evaluated as the policy declares.
Generation may end, for example, before the point where the policy evaluates.
Therefore, the serving system must tell apart an evaluation that completed, one that was never reached, and one that failed because required evidence was missing.

\textbf{Authorized release (R3).}
No token may be released before the evaluations that the policy requires for it have completed, and what is released must follow their outcome.
After a completed evaluation, the service releases the held output, the fixed refusal, or nothing further, according to the decision.
When an evaluation was not reached, the service applies the treatment that the policy declares for that case.
A request whose execution failed is never released as if the policy had allowed it.

Together, these requirements define \emph{mediation correctness}, the property LMSM provides and the subject of the first question above.
The second question concerns \emph{policy effectiveness}, which depends on each deployment.
Our evaluation reports both.

\section{LMSM Design}
\label{sec:design}

\begin{figure*}[t]
    \centering
    \includegraphics[
        width=0.94\textwidth
    ]{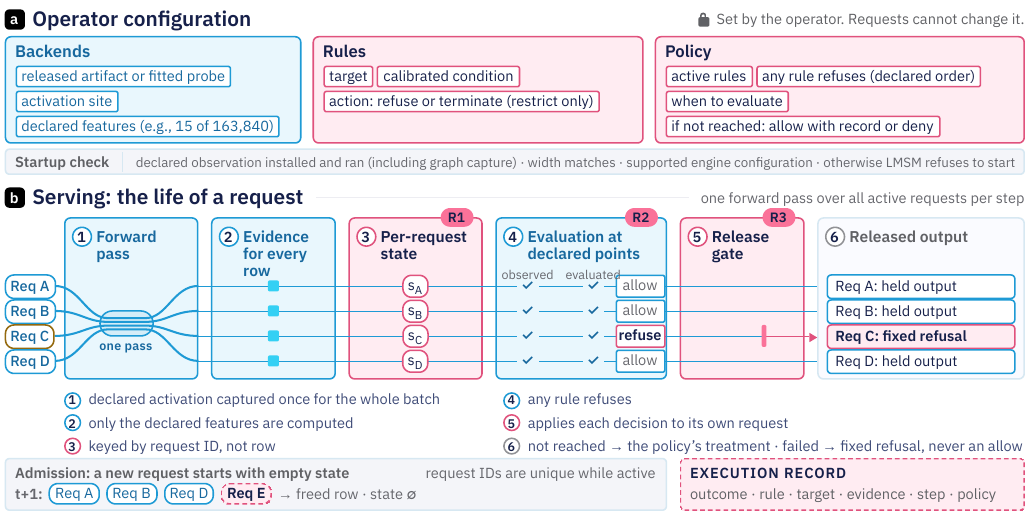}
    \caption{
        \textbf{The life of a request in LMSM.}
        (a) Before serving, the operator installs backends, rules, and a policy that declares the evidence it uses and when to evaluate it.
        A startup check confirms that the engine can deliver what the policy declares.
        (b) In each forward pass, LMSM captures the declared activations once for the whole batch and passes each row to the state of its request (R1).
        At the declared evaluation points, rules turn the request's evidence into a decision, and the request's execution record notes whether every declared observation and evaluation took place (R2).
        The release gate applies each decision to its own request and releases output only as the decision and the execution record allow (R3).
    }
    \label{fig:batch-aware}
\end{figure*}

The key idea of LMSM is to separate what differs between model-internal defenses from what they share.
These defenses differ in the interpretability method that produces evidence and in the rules that decide what this evidence allows.
They all require the same path from evidence to release, and LMSM implements this path once.
An interpretability method joins the path as a backend, and the operator's rules join it as a policy, each through a narrow interface.
Through the backend interface, a backend only turns the activations of each request into named evidence.
Through the policy declaration, a policy only states which evidence it uses, when to evaluate it, and how to restrict output.
Because the shared path relies on nothing else, an existing or future interpretability method becomes a runtime defense through a backend adapter and calibrated rules.
It then inherits all three duties unchanged.
LSM makes the same split in the kernel, and Table~\ref{tab:design-glance} sets each part of our design beside the LSM mechanism it adapts.
We first describe the backend interface and the policy declaration.
We then follow a request along the shared path in Figure~\ref{fig:batch-aware}.
The path keeps each request's state with that request (R1, Section~\ref{sec:policy-evaluator}), executes the declared policy (R2, Section~\ref{sec:execution}), and releases output only as the policy allows (R3, Section~\ref{sec:enforcement-gate}).
Finally, we show how existing and future methods plug into this path.

\begin{table}[t]
  \centering
  \caption{
    LSM and LMSM side by side.
  }
  \label{tab:design-glance}
  \footnotesize
  \setlength{\tabcolsep}{3pt}
  \renewcommand{\arraystretch}{1.10}
  \begin{tabularx}{\columnwidth}{
    >{\raggedright\arraybackslash}p{0.22\columnwidth}
    >{\raggedright\arraybackslash}X
    >{\raggedright\arraybackslash}X
  }
    \toprule
    \textbf{Design choice} & \tblswatch{LMSMBlue}\ \textbf{LSM in the kernel} & \tblswatch{LMSMGreen}\ \textbf{LMSM in LLM serving} \\
    \midrule
    Shared path & Fixed hooks with loadable security modules & One shared path with pluggable backends and policies \\
    \midrule
    Declarations & A module registers the hooks it uses & A policy declares the evidence and evaluations it uses \\
    \midrule
    Restrictive rules & Restrictive hooks deny but never grant & Rules only refuse or terminate \\
    \midrule
    Per-request state & Security fields in kernel objects & State kept with each request \\
    \midrule
    Fail-safe default~\cite{saltzer1975protection} & A hook error denies the access & An execution failure is never an allow \\
    \bottomrule
  \end{tabularx}
\end{table}

\subsection{Backend Interface}
\label{sec:backend-contract}

To let any interpretability method supply evidence, the backend interface asks very little of a backend.
Each backend reads the activations at the sites it declares, such as the input to a layer's multilayer perceptron (MLP), and reports named scalar signals for each active request.
These signals are the request's evidence, such as the value of an SAE or transcoder feature or the margin of a probe (its signed distance from the decision boundary).
The interface places two requirements on a backend and no others.
The evidence for a request must depend only on that request's activations, and it must stay aligned with the request it came from.
The backend alone decides how to compute its evidence, and neither the policy nor the shared path can tell a released artifact from a probe fitted by the operator.

Because rules name the evidence they use, a backend can compute only that evidence.
The transcoder policy in our evaluation, for example, reads 15 of the artifact's 163,840 features.
Rules bound to the same backend share this computation, and all active requests share one batched backend call.
When the artifact's activation function is elementwise, acting on each feature separately as ReLU and JumpReLU do, the selected features take the values that the full encoder would produce, up to floating-point rounding.
For such artifacts, the cost of protection depends on the policy rather than on the size of the artifact.

\subsection{Policy Declaration}
\label{sec:policy-design}

To let the shared path execute a policy without understanding its evidence, LMSM treats each policy as a declaration.
The declaration states which evidence the active rules read, when they are evaluated, and how they restrict output, and the shared path relies on these statements alone.
For the restriction on intrusion instructions in our running example, a rule's condition may test whether a group of transcoder features exceeds calibrated thresholds or whether the margin of a probe exceeds a threshold.
Every rule is restrictive, meaning that its action can only refuse or terminate the request and can never grant permission or cancel the action of another rule.

If any active rule triggers, the policy applies that rule's action.
When several rules trigger together, an order that the operator declares picks one, which keeps each decision and its attribution deterministic.
Because every rule is restrictive, adding a rule can only make a policy stricter.
Rules over several backends can also share one policy, and a small composition step merges their declarations while capture, execution records, and release remain unchanged.

\begin{figure}[t]
    \centering
    \includegraphics[width=\columnwidth]{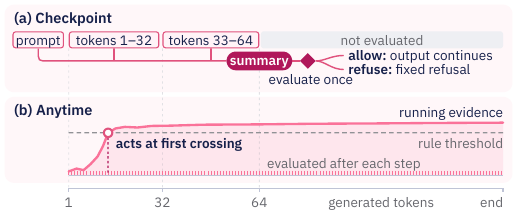}
    \caption{
        \textbf{Checkpoint and anytime schedules.}
        (a) A checkpoint schedule summarizes the prompt and two windows of generated tokens and evaluates its rules once, at token 64.
        (b) An anytime schedule updates its evidence after each decoding step and acts at the first threshold crossing.
    }
    \label{fig:schedules}
\end{figure}

The declaration also sets a schedule, which fixes the points at which the rules are evaluated.
A \textit{checkpoint schedule} evaluates once, after a fixed number of generated tokens such as 64.
An \textit{anytime schedule} evaluates after each decoding step and acts at the first threshold crossing.
The checkpoint schedule is cheaper but leaves later tokens unevaluated, while the anytime schedule covers every processed token at a higher cost (Figure~\ref{fig:schedules}).

Choosing between them is a matter of policy rather than of mediation.
Generation can also end before an evaluation point, as when a short answer finishes within its first 64 tokens.
The policy declares how LMSM treats such a request, either allowing it with a record of the missed evaluation or denying it.

\subsection{Request Association (R1)}
\label{sec:policy-evaluator}

To keep evidence with its request, LMSM keys all policy state by the request's identifier rather than by its row in the batch.
This state holds the evidence that the request has accumulated, any summaries that its schedule uses, and the decisions made so far.
Each forward pass captures the declared activations once for the whole batch, and LMSM passes each row to the state of the request that owns it.
LMSM creates empty state when the engine admits a request and deletes that state when the request finishes.
Without request keying, a request whose row moved would read another request's evidence.
Without this admission reset, a new request would inherit the evidence and decisions of the request it replaced.
A harmful response could then escape the refusal meant for it, and a benign one could receive another user's refusal.
Our evaluation measures both effects by removing each mechanism.
Because the state is keyed by request rather than by backend or rule, every backend and policy inherits this isolation without code of its own.

\subsection{Execution of the Declared Policy (R2)}
\label{sec:execution}

A policy is executed only if the observations and evaluations it declares actually take place, and a modern serving engine does not guarantee this.
Engines replay recorded CUDA graphs without running Python code, reuse cached prompt prefixes without recomputing them, and split long prompts across steps.
Each of these optimizations can change which positions a forward pass computes or whether an observer runs at all.
To close this gap, LMSM verifies execution twice, once before serving and again for each request.

Before serving, a startup check confirms that the engine can deliver what the policy declares.
It verifies that each declared activation is captured at its declared module with the declared width, that each capture ran while the engine started, and that the engine configuration is one that LMSM supports.
If any condition fails, LMSM refuses to start and names the condition.
The mediation correctness that LMSM provides covers exactly the configurations that pass this check.

During serving, LMSM keeps an execution record for each request.
For each position that the policy must observe, the record notes whether its observation arrived in the step that processed it.
For each evaluation point that the request reached, the evaluation itself records that it ran.
The record counts declared observations and evaluations without reading their values, and it requires no knowledge of what a backend computes.
When the request finishes, LMSM uses the record to assign it exactly one of three outcomes, checked in this order.
The outcome is \emph{execution failed} if a processed position lacks its observation or a reached evaluation did not run, \emph{not reached} if generation ended before an evaluation point, and \emph{completed} if every declared observation and evaluation took place.
As a result, an optimization that silently skips an observation or an evaluation leaves a visible execution failure.
In this way, complete mediation becomes a property that LMSM checks for every request rather than one it assumes.

\subsection{Authorized Release (R3)}
\label{sec:enforcement-gate}

To enforce each decision, LMSM lets output reach the requester only through a release gate.
The gate receives one decision and one outcome for each request and applies them to that request alone.
Output is released only after the evaluations that cover it have completed.
For a failed execution, the gate then releases the fixed refusal, and for a request that was not reached, it applies the treatment that the policy declares.
A completed request releases what its decision allows (Figure~\ref{fig:batch-aware}).
Tokens that the schedule never covers, such as those after a checkpoint, are released as the policy states.
A decision may also stop generation early to save compute.
LMSM still enforces each decision at the gate, and a request that keeps generating releases only what its decision allows.
The gate holds no backend logic, thresholds, or schedules, which is why a new method leaves it untouched.

\subsection{Adapting Existing and Future Methods}
\label{sec:audit-design}

To see what this separation provides, consider the three methods in our evaluation.
The fitted probes and the released transcoder on Qwen3-4B read the same site, the input to the MLP of layer 24, and swapping one for the other leaves all three duties unchanged.
The released SAE on Gemma-3-4B reads the residual stream after layer 22 instead, which required one new capture point, and it runs under the same checks in compiled vLLM.
Our evaluation confirms both cases.
Beyond these three, a future method fits the same interface as long as it turns each request's activations into named scalars, whether it reads one site or several.
Each new site adds a capture point, which the startup check and the execution record then cover like any other.
The only evidence that the interface excludes is evidence for one request that depends on other requests, as the first backend requirement states.
The execution record also states the request's outcome and, for each intervention, the rule, target, evidence, threshold, step, and policy version.
These records let the operator audit how a newly deployed method reaches its decisions.
In this way, adopting a new method requires work on that method alone.

\section{Implementation}
\label{sec:implementation}

We implement LMSM in vLLM and in Hugging Face Transformers.
Our main implementation runs inside compiled vLLM, which replays recorded CUDA graphs, and it keeps vLLM's default optimizations on.
It evaluates policies on the GPU without synchronizing with the host.
Eager vLLM and Transformers serve as simpler reference implementations, which we use to validate the compiled path.
We then describe the backends and policies we deploy and the configurations that our implementation supports.

\subsection{Mediation Inside Compiled vLLM}
\label{sec:runtime-integrations}

\begin{figure}[t]
    \centering
    \includegraphics[width=\columnwidth]{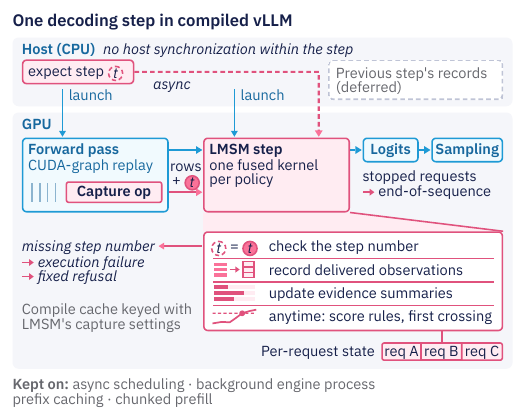}
    \caption{
        \textbf{One decoding step of LMSM in compiled vLLM.}
        The capture op runs inside the replayed CUDA graph, and one fused kernel per policy checks the step number and updates each request's state on the GPU before the logits are computed.
        The host sets the expected step number and processes the previous step's records after the next forward pass has started, which keeps the step free of host synchronization.
    }
    \label{fig:compiled-step}
\end{figure}

To observe activations in a model that replays recorded CUDA graphs without running Python code, LMSM places its capture inside the compiled model (Figure~\ref{fig:compiled-step}).
After the weights load, LMSM replaces the forward pass of the declared module with one that first calls a custom capture op.
For a module input, the op copies the activation rows of the batch into a fixed GPU buffer.
For the output of a decoder layer, it writes the residual stream that the layer passes on, which is the tensor that a Transformers hook would see.
In both cases, the op also writes the current step number into a second buffer.
Because both writes are ordinary GPU kernels on fixed addresses, CUDA-graph capture records them and each replay repeats them.
The evaluation on the GPU compares this number with the one the host expects.
If the capture did not run, the affected requests end in an execution failure and receive the fixed refusal.
LMSM also adds its capture configuration to vLLM's compile-cache key, which keeps vLLM from reusing a graph compiled without the capture.

To keep evaluation off the host, the policy state of each request lives on the GPU.
LMSM assigns a state slot to the request's identifier at admission and frees it when the request finishes, and a change of batch row never moves this state.
After each forward pass and before the logits are computed, one fused kernel per policy checks the step number, records the observations that arrived, and updates the evidence of each request.
The kernel also applies decisions on the GPU, where a decision that ends generation forces the end-of-sequence token without a host synchronization.
Only bookkeeping remains on the host, and LMSM writes the execution records after the next forward pass has started.
At startup, LMSM gathers the encoder weights of the features that the SAE and transcoder rules name, and it computes only those features on the GPU.
Encoders whose activation is not elementwise, such as TopK, run on the host, which computes all features before selecting the named ones.

Device evaluation also keeps vLLM's default optimizations on, and each of them requires one adjustment.
Under asynchronous scheduling, step indices come from per-request forward counters rather than from token counts on the host.
With vLLM's background engine process, the release gate runs inside the worker.
Under prefix caching and chunked prefill, LMSM observes the prompt at its last token, which vLLM computes even when the rest of the prompt comes from the cache.
Chunked prefill changes the numerics of the evidence much as a change of GPU kernels would, and our evaluation measures its effect.

\subsection{Reference Implementations}
\label{sec:reference-implementations}

To validate the compiled path, we keep two simpler implementations of the same design.
Eager vLLM reads the declared activation with a Python forward hook and evaluates policies on the host, and it shares the execution records and the release gate of the compiled path.
Transformers serves one request at a time and evaluates the policy on each candidate token before committing it.
It shares the backends and policies but keeps no execution records, and we use it as a numerical reference for activations and evidence.

\subsection{Deployed Backends and Policies}
\label{sec:backend-policy-implementation}

To cover both ways of provisioning a backend, we deploy fitted probes and two released artifacts, and Table~\ref{tab:backends} shows the parts specific to each method.
The probes are class-balanced logistic classifiers on Qwen3-4B~\cite{yang2025qwen3}, one per target, and each standardizes the activation with fitted means and scales and reports its signed margin as evidence.
The released transcoder for Qwen3-4B comes from the circuit-tracer project~\cite{hanna2025circuittracer}, and the released JumpReLU SAE for Gemma-3-4B~\cite{gemmateam2025gemma3} comes from Gemma Scope~2~\cite{mcdougall2025gemmascope2}.

\begin{table}[t]
  \centering
  \caption{
    Deployed backends.
  }
  \label{tab:backends}
  \footnotesize
  \setlength{\tabcolsep}{3pt}
  \renewcommand{\arraystretch}{1.10}
  \begin{tabularx}{\columnwidth}{
    >{\raggedright\arraybackslash}p{0.17\columnwidth}
    >{\raggedright\arraybackslash}p{0.25\columnwidth}
    >{\raggedright\arraybackslash}p{0.25\columnwidth}
    >{\raggedright\arraybackslash}X
  }
    \toprule
    \textbf{Backend} & \textbf{Model and site} & \textbf{Evidence} & \textbf{Method-specific parts} \\
    \midrule
    Fitted probes & Qwen3-4B, layer-24 MLP input & One margin per target, 15 targets & Probe adapter, fitted probes, thresholds \\
    \midrule
    Released transcoder & Qwen3-4B, layer-24 MLP input & 15 of 163,840 features in 6 rules & Encoder adapter, calibrated rules \\
    \midrule
    Released SAE & Gemma-3-4B, layer-22 residual stream & 15 of 262,144 features in 6 rules & Encoder adapter, calibrated rules, capture point \\
    \bottomrule
  \end{tabularx}
\end{table}

Each deployment is a configuration file that names its backend, its rules, and its policy, and switching the active rules requires no code change.
For the probes, we deploy a checkpoint policy and an anytime policy over 15 targets, each with its own fitted probes and thresholds.
The checkpoint policy evaluates once after generated token 64 and fuses evidence from the prompt and from tokens 1 to 32 and 33 to 64.
The anytime policy evaluates after each decoding step over a running summary of the generated prefix and acts at the first crossing.
Both apply the action of any active rule that triggers, and they break ties by the declared rule order.
Our implementation captures one activation per deployed policy and composes only backends that read that activation, as the probes and the transcoder do.

For release, the gate holds the whole response until generation ends, which lets a refusal discard everything the model produced.
The gate also adds each intervention to the request's execution record, with the rule, target, score, threshold, step, and policy version.
These records show which rule caused each refusal, and an operator can use them to keep or drop rules by target.

\subsection{Supported Configurations}
\label{sec:supported-configurations}

Our implementation supports a fixed set of engine configurations and refuses to start in any other, which makes the scope of its guarantees explicit.
It supports vLLM 0.19 with its V1 model runner, in compiled mode with full and piecewise CUDA graphs or in eager mode.
With device evaluation, it also supports asynchronous scheduling, the background engine process, prefix caching, and chunked prefill.
It does not support speculative decoding or tensor, pipeline, or data parallelism.
A capture point must be a module input or the output of a decoder layer whose residual handling we have verified to match Transformers, which currently covers Gemma~3.
Device evaluation covers the probe policies and the SAE and transcoder encoders whose activation is elementwise.
Other backends run on the host, which requires vLLM's synchronous settings.
Finally, vLLM's batch-invariant mode, which removes the dependence of evidence on batch composition, is available for Qwen3-4B but not for Gemma-3-4B.
Section~\ref{sec:discussion} discusses the limits of the design itself.

\section{Evaluation}
\label{sec:evaluation}

We evaluate LMSM through four research questions, the first two of which are the two questions of our security model.
\textbf{RQ1} asks whether LMSM applies each policy correctly under continuous batching and what fails without its mechanisms.
\textbf{RQ2} asks how much harmful output the deployed policies remove and how they compare with the alignment and external-guard layers.
\textbf{RQ3} asks whether an operator can deploy released interpretability artifacts and turn their rules on and off by target.
Finally, \textbf{RQ4} asks what mediation costs when vLLM keeps its optimizations on and what the alternatives would cost.

\subsection{Setup}
\label{sec:eval-setup}

\textbf{Deployments.}
We evaluate the three deployments of Table~\ref{tab:backends} in compiled vLLM on one NVIDIA H100 GPU with up to 32 concurrent requests.
For effectiveness, we use matched settings, which turn off asynchronous scheduling, the background engine process, prefix caching, and chunked prefill, as host evaluation requires.
We repeat the main results with vLLM's default optimizations, which the cost results use.
As a baseline, the Disabled deployment runs the same path as Checkpoint with its actions suppressed, which isolates the effect of enforcement and shows what the aligned model does on its own.
We measure effectiveness on Qwen3-4B and Gemma-3-4B and cost on Qwen3 models of up to 32 billion parameters.

\textbf{Benchmarks and judges.}
We report attack success rates (ASRs) on 602 HarmBench prompts~\cite{mazeika2024harmbench}, 313 StrongREJECT prompts~\cite{souly2024strongreject}, and 2,000 adversarial WildJailbreak prompts~\cite{jiang2024wildteaming}, judged by Llama Guard~3~\cite{inan2023llama} on the released response.
Similarly, we report false-refusal rates (FRRs) on the 250 safe XSTest prompts~\cite{rottger2024xstest}, judged by WildGuard~\cite{han2024wildguard}.
Both rates count judged responses rather than interventions.
The prompts and judge rules follow ThinkSafe~\cite{lee2026thinksafe}, which reports training-time defenses on the same model.
For generation, we use nucleus sampling with temperature 0.6 and up to 16,384 new tokens with reasoning on.
Throughout, each rate is the mean over three seeds, with the lowest and highest seed as subscripts in the tables.

\textbf{Fitting and test data.}
We fitted the probes and their thresholds only on WildJailbreak's training split, and none of its 2,000 evaluation prompts appears in that split, either verbatim or as a near duplicate.
Similarly, only 2 of the 250 XSTest prompts appear in the fitting data, and excluding them changes each FRR by less than 0.05 points.
Moreover, HarmBench contributed no fitting data.
Finally, both judges read the entire released response, including its reasoning trace, and the appendix describes the remaining settings.

\subsection{Mediation Correctness}
\label{sec:eval-mediation}

\begin{table}[t]
  \centering
  \caption{
    Mediation tests.
  }
  \label{tab:mediation-tests}
  \footnotesize
  \setlength{\tabcolsep}{4pt}
  \renewcommand{\arraystretch}{1.12}
  \providecommand{\cmark}{\textcolor{LMSMGreenDark}{\ensuremath{\boldsymbol{\checkmark}}}}
  \begin{tabularx}{\columnwidth}{
    >{\raggedright\arraybackslash}X
    >{\raggedleft\arraybackslash}p{0.12\columnwidth}
    >{\raggedright\arraybackslash}p{0.36\columnwidth}
  }
    \toprule
    \textbf{Test} & \textbf{Count} & \textbf{Result} \\
    \midrule
    \multicolumn{3}{l}{\textbf{R1}\enspace\emph{Request association}} \\
    Probe policies & 23,856 & \cmark\ No association error \\
    Released artifacts & 13,632 & \cmark\ No association error \\
    Duplicate pairs, batch-invariant & 5,964 & \cmark\ All identical \\
    \midrule
    \multicolumn{3}{l}{\textbf{R2}\enspace\emph{Execution of the declared policy}} \\
    Unsupported configurations & 11 & \cmark\ All refused at startup \\
    Capture removed after startup & 384 & \cmark\ All failed closed \\
    One observation dropped & 32 & \cmark\ All failed closed \\
    Device versus host evaluation & 128 & \cmark\ Same decisions \\
    Not reached on short answers & 5,704 & \cmark\ 0 mismatches \\
    \midrule
    \multicolumn{3}{l}{\textbf{R3}\enspace\emph{Authorized release}} \\
    Failed requests & 416 & \cmark\ Fixed refusal only \\
    Released-artifact policies & 7,668 & \cmark\ Released as decided \\
    \bottomrule
  \end{tabularx}
\end{table}

\begin{table*}[t]
  \centering
  \caption{
    Attack success and false refusals on Qwen3-4B (\%).
    Bold marks the best defense.
  }
  \label{tab:policy-effectiveness}
  \footnotesize
  \setlength{\tabcolsep}{3pt}
  \renewcommand{\arraystretch}{1.10}
  \providecommand{\rng}[2]{\textsubscript{\,#1--#2}}
  \begin{tabularx}{0.95\textwidth}{
    >{\raggedright\arraybackslash}X
    >{\centering\arraybackslash}p{0.08\textwidth}
    *{3}{>{\centering\arraybackslash}p{0.12\textwidth}}
    >{\centering\arraybackslash}p{0.15\textwidth}
    >{\centering\arraybackslash}p{0.11\textwidth}
  }
    \toprule
    & & \multicolumn{3}{c}{\textbf{Attack success rate} $\downarrow$} & \textbf{False-refusal rate} $\downarrow$ & \\
    \cmidrule(lr){3-5} \cmidrule(lr){6-6}
    \textbf{Configuration} & \textbf{Weights} & \textbf{HarmBench} & \textbf{StrongREJECT} & \textbf{WildJailbreak} & \textbf{XSTest} & \textbf{Mean} $\downarrow$ \\
    \midrule
    \multicolumn{7}{l}{\emph{LMSM on Qwen3-4B under matched settings (StrongREJECT with vLLM's default optimizations)}} \\
    \tblswatch{LMSMDisabled}\ Disabled & Unchanged & 39.37\rng{39.04}{39.87} & 6.60\rng{4.79}{8.31} & 43.23\rng{42.75}{43.60} & 0.80\rng{0.80}{0.80} & 22.50\rng{22.06}{22.91} \\
    \rowcolor{LMSMGreenTint}
    \tblswatch{LMSMCheckpoint}\ Checkpoint & Unchanged & \textbf{3.27}\rng{3.16}{3.49} & 1.38\rng{1.28}{1.60} & 6.72\rng{6.35}{7.10} & 3.07\rng{2.80}{3.60} & \textbf{3.61}\rng{3.57}{3.66} \\
    \rowcolor{LMSMGreenTint}
    \tblswatch{LMSMAnytime}\ Anytime & Unchanged & 5.54\rng{5.15}{5.81} & 2.34\rng{2.24}{2.56} & \textbf{6.25}\rng{5.80}{6.60} & 5.20\rng{5.20}{5.20} & 4.83\rng{4.60}{5.00} \\
    \midrule
    \multicolumn{7}{l}{\emph{Qwen3-4B with training-time defenses, as reported by ThinkSafe}} \\
    \tblswatch{LMSMDisabled}\ No safety training & Unchanged & 38.21 & 8.31 & 43.00 & 0.80 & 22.58 \\
    \rowcolor{LMSMBlueTint}
    \tblswatch{LMSMBlue}\ DirectRefusal~\cite{huang2025safetytax} & Fine-tuned & 33.06 & 3.19 & 36.20 & 32.00 & 26.11 \\
    \rowcolor{LMSMBlueTint}
    \tblswatch{LMSMBlue}\ SafeChain~\cite{jiang2025safechain} & Fine-tuned & 43.69 & 41.21 & 39.65 & 2.00 & 31.64 \\
    \rowcolor{LMSMBlueTint}
    \tblswatch{LMSMBlue}\ STAR-1~\cite{wang2026star1} & Fine-tuned & 33.72 & 5.75 & 35.15 & 6.80 & 20.36 \\
    \rowcolor{LMSMBlueTint}
    \tblswatch{LMSMBlue}\ SafePath~\cite{jeung2025safepath} & Fine-tuned & 37.71 & 7.35 & 42.45 & 1.60 & 22.28 \\
    \rowcolor{LMSMBlueTint}
    \tblswatch{LMSMBlue}\ SafeKey~\cite{zhou2025safekey} & Fine-tuned & 32.39 & 3.19 & 32.95 & \textbf{0.80} & 17.33 \\
    \rowcolor{LMSMBlueTint}
    \tblswatch{LMSMBlue}\ ThinkSafe~\cite{lee2026thinksafe} & Fine-tuned & 9.63 & \textbf{0.32} & 7.45 & 2.80 & 5.05 \\
    \bottomrule
  \end{tabularx}
\end{table*}

We first test whether LMSM performs the three duties in compiled vLLM, independently of how well any policy recognizes harmful output.
Table~\ref{tab:mediation-tests} summarizes the tests, which use the probe policies unless noted, and the appendix describes their workloads.

\textbf{Request association (R1).}
LMSM keeps each request's evidence, state, and decisions with that request.
To stress association, we submit each of 852 prompts twice and let continuous batching move requests between rows, reuse the rows of finished requests, and mix prompt processing with decoding.
Across 37,488 request executions on the three backends, no evidence, state, or decision reached another request.
In batch-invariant mode, both copies of each prompt also received the same decision in all 5,964 duplicate pairs (Table~\ref{tab:mediation-tests}).
Under default kernels, the two copies received different decisions in 0.8 to 6.3\% of the pairs under the probe policies, because the numerics of a forward pass depend on which requests share it.
Each of these decisions still relied on its own request's evidence.

\begin{figure}[t]
    \centering
    \includegraphics[width=\columnwidth]{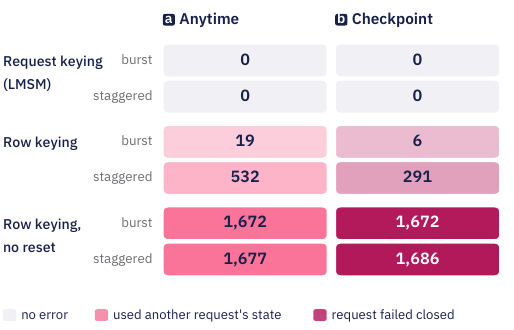}
    \caption{
        \textbf{Effect of removing request keying and the admission reset.}
        With the same detectors and rules, row keying lets state cross requests, and Anytime then decides on another request's state while Checkpoint fails the affected requests closed.
        Each tile counts requests out of the 1,704 in its run, in batch-invariant mode with host evaluation.
    }
    \label{fig:ablation}
\end{figure}

To see what request keying and the admission reset prevent, we then keep the detectors and rules unchanged, key the state by batch row instead of by request, and also remove the admission reset (Figure~\ref{fig:ablation}).
Anytime has no check that notices either change.
Without the reset, 904 of its 1,704 decisions change, including 878 HarmBench requests that LMSM refused and the row-keyed policy releases.
As a result, ordinary batching alone produces both outcomes in which the attacker of our threat model wins.
Decisions rely on another request's state, and the service releases responses that the policy prohibits.
In contrast, the execution records of the affected requests show a missing observation or an evaluation that did not run, and Checkpoint fails these requests closed instead of deciding on another request's state.
Therefore, request keying and the admission reset prevent both outcomes.
Even without them, a policy that verifies its own execution refuses the affected requests instead of releasing them silently.

\textbf{Execution and release (R2 and R3).}
LMSM executes the declared policy and releases only what it allows.
The startup check refused each unsupported configuration before serving, each request that lost its capture or an observation failed closed, and the requests around them kept their state.
Moreover, device evaluation reached the same decision and release as host evaluation for each compared request.
Finally, when answers end before token 64, the execution records mark exactly those requests as not reached, and the treatment that the policy declares decides their release.
Overall, release followed the recorded outcome in each test, and LMSM met all three duties in each configuration that passed its startup check, including vLLM's default optimizations.

\subsection{Policy Effectiveness}
\label{sec:eval-effectiveness}

Given that LMSM applies each policy correctly, we next measure how well the deployed policies stop harmful output and how often they refuse benign requests.

\begin{table}[t]
  \centering
  \caption{
    LMSM and a Llama Guard~3 output filter (\%).
  }
  \label{tab:filter}
  \footnotesize
  \setlength{\tabcolsep}{3pt}
  \renewcommand{\arraystretch}{1.10}
  \providecommand{\rng}[2]{\textsubscript{\,#1--#2}}
  \begin{tabularx}{\columnwidth}{
    >{\raggedright\arraybackslash}X
    >{\centering\arraybackslash}p{0.17\columnwidth}
    >{\centering\arraybackslash}p{0.21\columnwidth}
    >{\centering\arraybackslash}p{0.14\columnwidth}
    >{\centering\arraybackslash}p{0.12\columnwidth}
  }
    \toprule
    & \multicolumn{2}{c}{\textbf{Harmful releases} $\downarrow$} & \textbf{FRR} $\downarrow$ & \textbf{Time} $\downarrow$ \\
    \cmidrule(lr){2-3} \cmidrule(lr){4-4} \cmidrule(lr){5-5}
    \textbf{Defense} & \textbf{HarmBench} & \textbf{WildJailbreak} & \textbf{XSTest} & \textbf{added} \\
    \midrule
    \tblswatch{LMSMBlue}\ Llama Guard~3 & 3.3\rng{2.9}{3.7} & 8.9\rng{7.8}{9.7} & \textbf{1.1}\rng{0.8}{1.2} & 9.6 \\
    \rowcolor{LMSMGreenTint}
    \tblswatch{LMSMCheckpoint}\ Checkpoint & 1.7\rng{1.5}{1.8} & 4.6\rng{4.4}{4.8} & 3.1\rng{2.8}{3.6} & \textbf{3.7} \\
    \rowcolor{LMSMGreenTint}
    \tblswatch{LMSMCheckpoint}\tblswatch{LMSMBlue}\ Combined & \textbf{0.5}\rng{0.3}{0.8} & \textbf{1.9}\rng{1.6}{2.1} & 3.3\rng{3.2}{3.6} & 10.2 \\
    \bottomrule
  \end{tabularx}
\end{table}

\textbf{Harmful output and false refusals.}
Both probe policies remove most harmful output at a small cost in false refusals.
As shown in Table~\ref{tab:policy-effectiveness}, Checkpoint reduces the HarmBench ASR from 39.4\% to 3.3\% and the WildJailbreak ASR by a factor of six under matched settings, while the XSTest FRR rises from 0.8\% to 3.1\%.
Moreover, each LMSM rate varies by at most 0.8 points across seeds.
With vLLM's default optimizations, the HarmBench, WildJailbreak, and XSTest rates change by at most 0.7 points.
Of the two schedules, Anytime acts earlier, at a median of 14 generated tokens instead of 64 (Figure~\ref{fig:schedules}).

\textbf{Beyond the development data.}
The reduction holds on prompts unseen during development and under a second judge, and ordinary requests are rarely affected.
With vLLM's default optimizations, Checkpoint reduces the ASR on the 224 HarmBench prompts whose behaviors never appeared during development from 36.2\% to 2.2\%.
The official HarmBench classifier confirms this reduction, rating 50.1\% of Disabled and 8.4\% of Checkpoint responses as harmful.
Meanwhile, Checkpoint intervenes on only 2.2\% of 3,042 ordinary requests.
Almost all are open-ended No Robots requests, and none is a GSM8K or HumanEval+ task.

\textbf{Comparison with training-time defenses.}
To compare with the alignment layer, the lower block of Table~\ref{tab:policy-effectiveness} shows the same model without safety training and with six training-time defenses, as reported by ThinkSafe.
Without safety training, this model scores within 2 points of our Disabled baseline on each benchmark, which supports reading the two blocks together.
Both LMSM policies reach a lower mean of the four rates than these defenses, a mean that ThinkSafe calls its safety average.
However, we did not rerun the defenses, and their XSTest generation differs from ours.
Since LMSM leaves the weights unchanged, it can also run on a model that such a defense has trained, with probes fitted to that model.

\textbf{Comparison with an output guard.}
Checkpoint also removes more harmful output than a Llama Guard~3 output filter, a widely used external guard (Table~\ref{tab:filter}).
To keep the filter from judging itself, WildGuard judges both.
Checkpoint releases harmful content about half as often as the filter and adds less than half of the filter's time, but its FRR is 2 points higher.
Since Checkpoint acts during generation and the filter acts on the released response, a deployment can also run both.
Together, they release 30 to 40\% of the harmful content that Checkpoint releases alone, while the FRR rises by only 0.3 points.
Running both takes about as long as the filter alone, because the filter reads only the short fixed refusal for each request that Checkpoint refuses.
Such results show that, with a mediation path, the model-internal layer is competitive with the alignment and external-guard layers and complements them.

\subsection{Released Artifacts and Rule Control}
\label{sec:eval-artifacts}

\begin{figure}[t]
    \centering
    \includegraphics[width=\columnwidth]{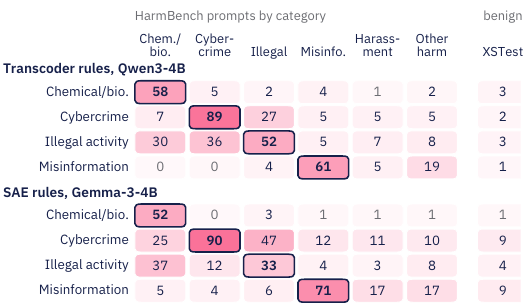}
    \caption{
        \textbf{Firing rates of the released-artifact category rules.}
        Seven of the eight rules fire most often on prompts of their own target, which is outlined, with overlaps where the categories themselves overlap.
        Each value is the share of prompts on which the rule alone fires, as a mean of three seeds.
        Qwen3-4B runs under matched settings and Gemma-3-4B under vLLM's default optimizations.
    }
    \label{fig:rules}
\end{figure}

\begin{figure*}[t]
    \centering
    \includegraphics[width=0.95\textwidth]{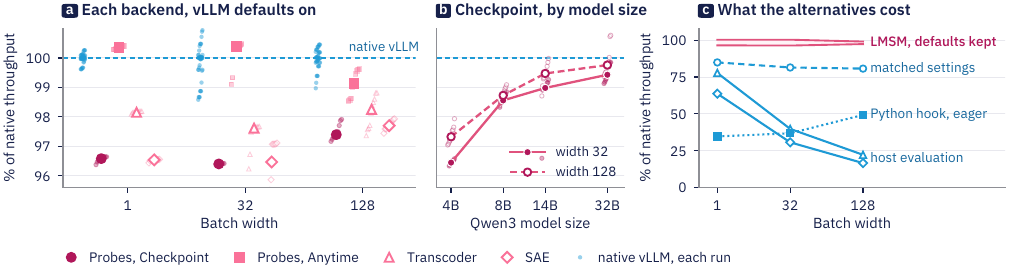}
    \caption{
        \textbf{Serving cost relative to native vLLM.}
        Each backend keeps at least 96.4\% of native throughput, the cost of a Checkpoint policy decreases as the model size increases, and the integrations that other designs require cost far more.
        (a) Each backend with vLLM's default optimizations on.
        (b) A Checkpoint policy with the released structure on larger Qwen3 models.
        (c) On Qwen3-4B, the range of panel (a) together with native vLLM under matched settings, host evaluation of the transcoder's features, and the Checkpoint policy with a Python forward hook, which requires eager execution.
        The diamonds show host evaluation of the SAE's features on Gemma-3-4B.
        Large markers are medians of ten measured repetitions, which the small markers show, and the \lmsmnativedots{} dots show the repetitions of native vLLM, whose spread covers the values slightly above 100\%.
    }
    \label{fig:cost}
\end{figure*}

We next ask whether an operator can instead deploy released interpretability artifacts and control them rule by rule.

\textbf{Selecting rules by target.}
The transcoder and SAE policies run on the same path.
Each artifact has four category rules, one per harm category, and two copyright rules, each reading a few named features.
Neither artifact was trained for safety, and their thresholds were set once, before our evaluation (see the appendix).
Even so, seven of the eight category rules fire most often on prompts of their own target, with overlaps where the categories themselves overlap, such as cybercrime and illegal activity (Figure~\ref{fig:rules}).
As a result, an operator can turn rules on and off by target, and the execution records attribute each refusal to its rule.
In the deployed runs, the records name a rule for each refusal and for no released response.

\textbf{Testing rules before deployment.}
Each rule depends only on the largest value that each of its features reaches, and the gate holds each response until generation ends.
Therefore, one Disabled run shows what any subset of rules would release (see the appendix).
Replaying all six rules in this way agrees with the deployed policies to within a point.
As a result, an operator can test each rule alone before deploying it.
Each of the transcoder's four category rules alone reduces the ASR on its own category by half or more, while the FRR rises from 0.8\% to at most 3.7\%.
The transcoder's cybercrime rule, for example, reduces the ASR on cybercrime prompts from 46.5\% to 1.0\%, while its FRR is 2.7\%.
The SAE's chemical and biological rule reduces the ASR on its category from 46.3\% to 25.9\% and increases the FRR by 0.9 points.
In contrast, the SAE's cybercrime and misinformation rules also fire on 9\% of XSTest prompts (Figure~\ref{fig:rules}).

\textbf{A target beyond safety training.}
A deployment may also forbid requests beyond the scope of safety training, such as requests for the text of a book or a song.
Each artifact has one rule for books and one for song lyrics.
On 74 held-out HarmBench requests of this kind, each rule fires on 92 to 96\% of the requests for its own kind of text and on at most 2\% of the others.
Together, the two rules of each artifact refuse 94 to 97\% of these requests.
On XSTest, three of the four rules increase the FRR by at most 0.4 points.
The transcoder's book rule increases it by 7 points, and a deployment that does not need this rule can leave it off.
Since both models rarely reproduce these texts word for word (see the appendix), the rules serve mainly to refuse requests that the deployment forbids.

Such results show that released artifacts suit targets that an existing representation already captures, such as copyrighted text, while an operator fits probes for the others.
Since the same mediation path runs probe and transcoder rules as one policy, a deployment can combine both.

\subsection{Serving Cost}
\label{sec:eval-cost}

Finally, we compare each deployment with native vLLM serving the same model in the same session, with each request generating exactly 256 tokens under greedy decoding at batch widths of 1, 32, and 128.
LMSM evaluates its policy at each declared point without acting, which keeps the work identical across runs.
We report the median throughput over ten measured repetitions in two rounds of alternating order (see the appendix).

\textbf{Default optimizations.}
With vLLM's default optimizations on, each backend keeps at least 96.4\% of native throughput.
As shown in Figure~\ref{fig:cost}a, this holds at each batch width, and under matched settings the probe policies keep 97.9 to 99.5\%.

\textbf{Cost of the alternatives.}
The integrations that other designs require cost far more than LMSM does (Figure~\ref{fig:cost}c).
Matched settings, which host evaluation requires, already give up 15 to 19\% of native throughput on Qwen3-4B.
Evaluating the transcoder's features on the host then keeps only 22\% at width 128, and the SAE's only 17\%, because each referenced feature value forces a synchronization.
Similarly, a Python forward hook requires eager execution, and the same Checkpoint policy then keeps only 35 to 49\% of native throughput.
LMSM avoids both costs by capturing activations inside the compiled graph and evaluating its rules on the device.

\textbf{Larger models.}
The cost of a Checkpoint policy decreases as the model size increases.
On Qwen3 models with 8, 14, and 32 billion parameters, a Checkpoint policy with the same structure keeps 98.6 to 99.4\% of native throughput at width 32, compared with 96.4\% on Qwen3-4B (Figure~\ref{fig:cost}b).
The cost decreases because the policy's work per step remains nearly constant while the cost of the forward pass increases with model size.
In these runs, the policy's parameters are synthetic, because only its cost is measured.
Such results show that LMSM keeps protection within a few percent of native serving while vLLM keeps its optimizations on.

\section{Related Work}
\label{sec:related-work}

Section~\ref{sec:motivation} compares LMSM with the other layers of defense, and this section compares it with work that shares its evidence, its place in the serving system, or its architecture.

\textbf{Model-internal defenses.}
Interpretability research now releases features at scale, such as the SAEs of Gemma Scope~2~\cite{mcdougall2025gemmascope2} and the transcoders built for circuit tracing~\cite{hanna2025circuittracer}.
Defenses built on such signals pair a detector with an action of their own.
HiddenDetect~\cite{jiang2025hiddendetect} marks jailbreak prompts, JBShield~\cite{zhang2025jbshield} and CAST~\cite{lee2025cast} steer activations toward a safe answer, NExT-Guard~\cite{fang2026nextguard} stops a response stream, and circuit breakers~\cite{zou2024circuit} train harmful representations away.
Production systems also run streaming classifiers~\cite{cunningham2026constitutional} and activation probes~\cite{kramar2026probes} at scale.
However, each of these methods leaves request association and release to its deployment.
Their detectors can serve as LMSM backends.

\textbf{Serving engines and external guards.}
Serving engines such as vLLM~\cite{kwon2023vllm} batch requests continuously and share KV-cache memory between them.
This sharing creates risks across requests, since cached prefixes can leak another tenant's prompt~\cite{wu2025promptpeek} and backdoored models can leak data between requests in the same batch~\cite{kuchler2026withinbatch}.
Observability tools such as vLLM Hook~\cite{ko2026vllmhook,fang2026mia}, DMI-Lib~\cite{yu2026dmilib}, and vLLM-Lens~\cite{cooney2026vllmlens} are not defenses, but they attribute each captured activation to its request.
They provide the per-request capture that LMSM also requires and leave the policy, its execution records, and release to analysis code.
External guards such as NeMo Guardrails~\cite{rebedea2023nemo} and LlamaFirewall~\cite{chennabasappa2025llamafirewall} instead gate the release of text.
LMSM moves this gate inside the engine and binds it to internal evidence, and our evaluation compares it with a Llama Guard~\cite{inan2023llama} filter.

\textbf{Reference monitors, from operating systems to agents.}
LMSM follows the reference-monitor approach that LSM brought to the kernel, and earlier monitors differ mainly in where they sit.
Inlined monitors~\cite{erlingsson2000irm} embed their checks in the program they protect, system-call interposition~\cite{provos2003systrace} mediates a process from outside, and virtual-machine introspection~\cite{garfinkel2003vmi} observes a guest from the hypervisor.
Recent work~\cite{christodorescu2026agentsecurity,christodorescu2025sok,beurerkellner2025designpatterns} applies these principles to LLM agents and treats the model as untrusted.
This work mediates the model's inputs and the agent's actions through isolation~\cite{wu2025isolategpt}, capabilities~\cite{debenedetti2025camel}, information-flow control~\cite{costa2025fides}, privilege control~\cite{shi2025progent}, runtime rules~\cite{wang2026agentspec}, structured queries~\cite{chen2025struq}, and operating-system abstractions~\cite{zou2026agentkernel}.
LMSM applies the same principles one layer down, to the evidence that the model computes.
As a result, an agent framework can enforce its tool policies over a model that LMSM serves, and each mediates what the other cannot observe.

\section{Discussion}
\label{sec:discussion}

LMSM splits a runtime defense into two parts.
Backends and rules detect harm and can change as interpretability research advances, while the mediation path that applies them remains the same.
This split opens four opportunities, which we discuss before two open problems.

\textbf{From interpretability research to deployment.}
Most interpretability methods are evaluated offline on fixed datasets, and deploying one has meant building request handling and enforcement around it.
With LMSM, a new method requires only a backend adapter and calibrated rules, as the released transcoder and SAE in our evaluation show.
Once a method is deployed, its execution records show which rules fire on real requests and at which step.
Such feedback is rare in offline evaluations, and it can guide the choice of features and thresholds for the next version.
Moreover, our mediation tests apply unchanged to any new backend and can check it before deployment.

\textbf{Richer actions.}
Each of our rules refuses a request or stops its generation.
As a result, adding a rule can only make a policy stricter.
However, the mediation path could also host actions that change the answer, such as steering the model toward a safe answer~\cite{lee2025cast}, removing a harmful span, or regenerating from an earlier step.
These actions would reuse the way LMSM ties state to requests, checks evaluations, and releases output.
Unlike a refusal, steering and regeneration change the activations that other rules read.
Steering could then hide the evidence that a refusal rule relies on.
A policy that combines the two would have to declare which activations each rule reads, such as those captured before the steering takes effect.

\textbf{Coverage and streaming.}
A Checkpoint policy evaluates its rules once, at token 64, and an answer that ends earlier never reaches that point.
Instead of releasing such an answer silently, LMSM records the request as not reached and applies the treatment that the policy declares, such as releasing the answer with a record or refusing it.
As a result, operators can match the schedule to the service, with a checkpoint for long answers and an anytime schedule, which evaluates after each step, for short ones.
Streaming fits the same design, because the gate has to hold a token only until the evaluations that cover it finish.
A streaming policy accepts that a later refusal cannot withdraw sent tokens.

\textbf{Beyond jailbreaks.}
Nothing in the mediation path depends on the kind of harm that a backend detects.
For example, probes on activations can detect when a model drifts from its user's task after reading injected instructions~\cite{abdelnabi2025drift}.
As an LMSM backend, such a probe could stop an agent's response before the injected instructions reach its tools.
Similarly, the released artifacts include copyright rules, which a service could turn on with its safety rules.

\textbf{Open problems.}
Two limits of the design itself point to research that LMSM makes easier.
First, LMSM enforces what its backends detect, and an adaptive attacker can still craft inputs that evade a backend~\cite{bailey2026obfuscated}.
Because backends share one mediation path, operators can deploy several of them together, as we did with the probes and the transcoder rules, and test them with the same adaptive attacks.
However, whether such combinations increase the cost of these attacks is still unknown.
Second, LMSM trusts the operator and the components that the operator installs.
To narrow this trust, the gate and its records could run in trusted hardware~\cite{tramer2019slalom}, which would also let a provider show its users which policy each response passed.

\section{Conclusion}
\label{sec:conclusion}

This paper presents LMSM, which adapts the design of Linux Security Modules to LLM serving and implements, in one shared mediation path, the duties that turn model-internal signals into a defense.
LMSM's backends turn model internals into evidence, its policies decide what that evidence allows, and each request records whether the declared evaluations took place.
In compiled vLLM with continuous batching, LMSM meets all three duties, hosts fitted probes, a released transcoder, and a released SAE on the same path, and keeps at least 96\% of native throughput with vLLM's optimizations on.
Moreover, the probe policies remove most harmful output at a small cost in false refusals.
Together, these results show that existing and future interpretability methods can become runtime defenses without rebuilding the serving path.
Our code and evaluation scripts are available at \codeurl{https://github.com/xiuyuz/LMSM}.

\bibliographystyle{IEEEtran}
\bibliography{references_verified}

@inproceedings{
zhang2026alphaalign,
title={{AlphaAlign}: Incentivizing Safety Alignment with Extremely Simplified Reinforcement Learning},
author={Yi Zhang and An Zhang and XiuYu Zhang and Leheng Sheng and Yuxin Chen and Zhenkai Liang and Xiang Wang},
booktitle={The Fourteenth International Conference on Learning Representations},
year={2026},
url={https://openreview.net/forum?id=2XNb1JUKW3}
}

@misc{zou2023universal,
      title={Universal and Transferable Adversarial Attacks on Aligned Language Models}, 
      author={Andy Zou and Zifan Wang and Nicholas Carlini and Milad Nasr and J. Zico Kolter and Matt Fredrikson},
      year={2023},
      eprint-={2307.15043},
      archivePrefix={arXiv},
      primaryClass={cs.CL},
      url={https://arxiv.org/abs/2307.15043}, 
}

@misc{wallace2024instruction,
      title={The Instruction Hierarchy: Training {LLMs} to Prioritize Privileged Instructions}, 
      author={Eric Wallace and Kai Xiao and Reimar Leike and Lilian Weng and Johannes Heidecke and Alex Beutel},
      year={2024},
      eprint={2404.13208},
      archivePrefix={arXiv},
      primaryClass={cs.CR},
      url={https://arxiv.org/abs/2404.13208}, 
}

@inproceedings{ouyang2022instructgpt,
author = {Ouyang, Long and Wu, Jeff and Jiang, Xu and Almeida, Diogo and Wainwright, Carroll L. and Mishkin, Pamela and Zhang, Chong and Agarwal, Sandhini and Slama, Katarina and Ray, Alex and Schulman, John and Hilton, Jacob and Kelton, Fraser and Miller, Luke and Simens, Maddie and Askell, Amanda and Welinder, Peter and Christiano, Paul and Leike, Jan and Lowe, Ryan},
title = {Training language models to follow instructions with human feedback},
year = {2022},
isbn = {9781713871088},
publisher = {Curran Associates Inc.},
address = {Red Hook, NY, USA},
booktitle = {Proceedings of the 36th International Conference on Neural Information Processing Systems},
articleno = {2011},
numpages = {15},
location = {New Orleans, LA, USA},
series = {NIPS '22}
}

@misc{inan2023llama,
      title={{Llama Guard}: {LLM}-based Input-Output Safeguard for Human-{AI} Conversations}, 
      author={Hakan Inan and Kartikeya Upasani and Jianfeng Chi and Rashi Rungta and Krithika Iyer and Yuning Mao and Michael Tontchev and Qing Hu and Brian Fuller and Davide Testuggine and Madian Khabsa},
      year={2023},
      eprint={2312.06674},
      archivePrefix={arXiv},
      primaryClass={cs.CL},
      url={https://arxiv.org/abs/2312.06674}, 
}

@inproceedings{jiang2025hiddendetect,
    title = "{H}idden{D}etect: Detecting Jailbreak Attacks against Multimodal Large Language Models via Monitoring Hidden States",
    author = "Jiang, Yilei  and
      Gao, Xinyan  and
      Peng, Tianshuo  and
      Tan, Yingshui  and
      Zhu, Xiaoyong  and
      Zheng, Bo  and
      Yue, Xiangyu",
    editor = "Che, Wanxiang  and
      Nabende, Joyce  and
      Shutova, Ekaterina  and
      Pilehvar, Mohammad Taher",
    booktitle = "Proceedings of the 63rd Annual Meeting of the Association for Computational Linguistics (Volume 1: Long Papers)",
    month = jul,
    year = "2025",
    address = "Vienna, Austria",
    publisher = "Association for Computational Linguistics",
    url = "https://aclanthology.org/2025.acl-long.724/",
    pages = "14880--14893",
    ISBN = "979-8-89176-251-0"
}

@inproceedings{
huben2024sparse,
title={Sparse Autoencoders Find Highly Interpretable Features in Language Models},
author={Hoagy Cunningham and Aidan Ewart and Logan Riggs and Robert Huben and Lee Sharkey},
booktitle={The Twelfth International Conference on Learning Representations},
year={2024},
url={https://openreview.net/forum?id=F76bwRSLeK}
}

@inproceedings{
dunefsky2024transcoders,
title={Transcoders find interpretable {LLM} feature circuits},
author={Jacob Dunefsky and Philippe Chlenski and Neel Nanda},
booktitle={The Thirty-eighth Annual Conference on Neural Information Processing Systems},
year={2024},
url={https://openreview.net/forum?id=J6zHcScAo0}
}

@inproceedings{arditi2024refusal,
    title={Refusal in Language Models Is Mediated by a Single Direction},
    author={Andy Arditi and Oscar Obeso and Aaquib Syed and Daniel Paleka and Nina Panickssery and Wes Gurnee and Neel Nanda},
    booktitle={The Thirty-eighth Annual Conference on Neural Information Processing Systems},
    year={2024},
    url={https://openreview.net/forum?id=pH3XAQME6c}
}

@inproceedings{zou2024circuit,
title={Improving Alignment and Robustness with Circuit Breakers},
author={Andy Zou and Long Phan and Justin Wang and Derek Duenas and Maxwell Lin and Maksym Andriushchenko and Rowan Wang and J Zico Kolter and Matt Fredrikson and Dan Hendrycks},
booktitle={The Thirty-eighth Annual Conference on Neural Information Processing Systems},
year={2024},
url={https://openreview.net/forum?id=IbIB8SBKFV}
}

@inproceedings{zhang2025jbshield,
author = {Zhang, Shenyi and Zhai, Yuchen and Guo, Keyan and Hu, Hongxin and Guo, Shengnan and Fang, Zheng and Zhao, Lingchen and Shen, Chao and Wang, Cong and Wang, Qian},
title = {{JBShield}: Defending Large Language Models from Jailbreak Attacks through Activated Concept Analysis and Manipulation},
booktitle = {34th USENIX Security Symposium (USENIX Security 25)},
year = {2025}
}

@techreport{anderson1972reference,
  title       = {Computer Security Technology Planning Study},
  author      = {Anderson, James P.},
  institution = {Air Force Electronic Systems Division, Hanscom AFB},
  number      = {ESD-TR-73-51, Vol.~II},
  year        = {1972},
  month       = oct,
  address     = {Bedford, MA},
  note_        = {Available at \url{https://csrc.nist.gov/csrc/media/publications/conference-paper/1998/10/08/proceedings-of-the-21st-nissc-1998/documents/early-cs-papers/ande72.pdf}}
}

@ARTICLE{saltzer1975protection,
  author={Saltzer, Jerome H. and Schroeder, Michael D.},
  journal={Proceedings of the IEEE}, 
  title={The protection of information in computer systems}, 
  year={1975},
  volume={63},
  number={9},
  pages={1278-1308},
  doi={10.1109/PROC.1975.9939}}

@inproceedings {wright2002lsm,
    author = {Chris Wright and Crispin Cowan and Stephen Smalley and James Morris and Greg Kroah-Hartman},
    title = {{Linux} Security Modules: General Security Support for the {Linux} Kernel},
    booktitle = {11th USENIX Security Symposium (USENIX Security 02)},
    year = {2002},
    address = {San Francisco, CA},
    url- = {https://www.usenix.org/conference/11th-usenix-security-symposium/linux-security-modules-general-security-support-linux},
    publisher = {USENIX Association},
    month = aug
}

@inproceedings {loscocco2001selinux,
author = {Peter Loscocco and Stephen Smalley},
title = {Integrating Flexible Support for Security Policies into the {Linux} Operating System},
booktitle = {2001 USENIX Annual Technical Conference (USENIX ATC 01)},
year = {2001},
address = {Boston, MA},
url- = {https://www.usenix.org/conference/2001-usenix-annual-technical-conference/integrating-flexible-support-security-policies},
publisher = {USENIX Association},
month = jun
}

@misc{apparmorAppArmor,
	author = {{AppArmor}},
	title = {{AppArmor} Documentation},
	howpublished = {\url{https://apparmor.net/}},
	note = {accessed 2026-11-17}
}

@inproceedings{kwon2023vllm,
author = {Kwon, Woosuk and Li, Zhuohan and Zhuang, Siyuan and Sheng, Ying and Zheng, Lianmin and Yu, Cody Hao and Gonzalez, Joseph E. and Zhang, Hao and Stoica, Ion},
title = {Efficient Memory Management for Large Language Model Serving with {PagedAttention}},
year = {2023},
isbn = {9798400702297},
publisher = {Association for Computing Machinery},
address = {New York, NY, USA},
url = {https://doi.org/10.1145/3600006.3613165},
doi = {10.1145/3600006.3613165},
booktitle = {Proceedings of the 29th Symposium on Operating Systems Principles},
pages = {611--626},
numpages = {16},
location = {Koblenz, Germany},
series = {SOSP '23}
}

@inproceedings{
bailey2026obfuscated,
title={Obfuscated Activations Bypass {LLM} Latent-Space Defenses},
author={Luke Bailey and Alex Serrano and Abhay Sheshadri and Mikhail Seleznyov and Jordan Taylor and Erik Jenner and Jacob Hilton and Stephen Casper and Carlos Guestrin and Scott Emmons},
booktitle={The Fourteenth International Conference on Learning Representations},
year={2026},
url={https://openreview.net/forum?id=ktGmDGoWnB}
}

@inproceedings{rebedea2023nemo,
    title = "{NeMo Guardrails}: A Toolkit for Controllable and Safe {LLM} Applications with Programmable Rails",
    author = "Rebedea, Traian  and
      Dinu, Razvan  and
      Sreedhar, Makesh Narsimhan  and
      Parisien, Christopher  and
      Cohen, Jonathan",
    editor = "Feng, Yansong  and
      Lefever, Els",
    booktitle = "Proceedings of the 2023 Conference on Empirical Methods in Natural Language Processing: System Demonstrations",
    month = dec,
    year = "2023",
    address = "Singapore",
    publisher = "Association for Computational Linguistics",
    url = "https://aclanthology.org/2023.emnlp-demo.40/",
    doi = "10.18653/v1/2023.emnlp-demo.40",
    pages = "431--445",
}

@inproceedings{rafailov2023direct,
author = {Rafailov, Rafael and Sharma, Archit and Mitchell, Eric and Ermon, Stefano and Manning, Christopher D. and Finn, Chelsea},
title = {Direct Preference Optimization: Your Language Model is Secretly a Reward Model},
year = {2023},
publisher = {Curran Associates Inc.},
address = {Red Hook, NY, USA},
booktitle = {Proceedings of the 37th International Conference on Neural Information Processing Systems},
articleno = {2338},
numpages = {14},
location = {New Orleans, LA, USA},
series = {NIPS '23}
}

@misc{bai2022constitutional,
      title={Constitutional {AI}: Harmlessness from {AI} Feedback}, 
      author={Yuntao Bai and Saurav Kadavath and Sandipan Kundu and Amanda Askell and Jackson Kernion and Andy Jones and Anna Chen and Anna Goldie and Azalia Mirhoseini and Cameron McKinnon and Carol Chen and Catherine Olsson and Christopher Olah and Danny Hernandez and Dawn Drain and Deep Ganguli and Dustin Li and Eli Tran-Johnson and Ethan Perez and Jamie Kerr and Jared Mueller and Jeffrey Ladish and Joshua Landau and Kamal Ndousse and Kamile Lukosuite and Liane Lovitt and Michael Sellitto and Nelson Elhage and Nicholas Schiefer and Noemi Mercado and Nova DasSarma and Robert Lasenby and Robin Larson and Sam Ringer and Scott Johnston and Shauna Kravec and Sheer {El Showk} and Stanislav Fort and Tamera Lanham and Timothy Telleen-Lawton and Tom Conerly and Tom Henighan and Tristan Hume and Samuel R. Bowman and Zac Hatfield-Dodds and Ben Mann and Dario Amodei and Nicholas Joseph and Sam McCandlish and Tom Brown and Jared Kaplan},
      year={2022},
      eprint-={2212.08073},
      archivePrefix={arXiv},
      primaryClass={cs.CL},
      url={https://arxiv.org/abs/2212.08073}, 
}

@inproceedings{mazeika2024harmbench,
author = {Mazeika, Mantas and Phan, Long and Yin, Xuwang and Zou, Andy and Wang, Zifan and Mu, Norman and Sakhaee, Elham and Li, Nathaniel and Basart, Steven and Li, Bo and Forsyth, David and Hendrycks, Dan},
title = {{HarmBench}: A Standardized Evaluation Framework for Automated Red Teaming and Robust Refusal},
booktitle = {Proceedings of the 41st International Conference on Machine Learning (ICML)},
year = {2024},
url = {https://proceedings.mlr.press/v235/mazeika24a.html}
}

@inproceedings{
jiang2024wildteaming,
title={{WildTeaming} at Scale: From In-the-Wild Jailbreaks to (Adversarially) Safer Language Models},
author={Liwei Jiang and Kavel Rao and Seungju Han and Allyson Ettinger and Faeze Brahman and Sachin Kumar and Niloofar Mireshghallah and Ximing Lu and Maarten Sap and Yejin Choi and Nouha Dziri},
booktitle={The Thirty-eighth Annual Conference on Neural Information Processing Systems},
year={2024},
url={https://openreview.net/forum?id=n5R6TvBVcX}
}

@inproceedings{
han2024wildguard,
title={{WildGuard}: Open One-stop Moderation Tools for Safety Risks, Jailbreaks, and Refusals of {LLM}s},
author={Seungju Han and Kavel Rao and Allyson Ettinger and Liwei Jiang and Bill Yuchen Lin and Nathan Lambert and Yejin Choi and Nouha Dziri},
booktitle={The Thirty-eighth Annual Conference on Neural Information Processing Systems Datasets and Benchmarks Track},
year={2024},
url={https://arxiv.org/abs/2406.18495}
}

@inproceedings{lee2026thinksafe,
      title={{ThinkSafe}: Self-Generated Safety Alignment for Reasoning Models},
      author={Seanie Lee and Sangwoo Park and Yumin Choi and Gyeongman Kim and Minki Kang and Jihun Yun and Sung Ju Hwang and Dongmin Park and Jongho Park},
      booktitle={The Fortieth Annual Conference on Neural Information Processing Systems},
      year={2026},
      url={https://arxiv.org/abs/2601.23143v5},
}

@inproceedings{rottger2024xstest,
    title = "{XST}est: A Test Suite for Identifying Exaggerated Safety Behaviours in Large Language Models",
    author = {R{\"o}ttger, Paul  and
      Kirk, Hannah  and
      Vidgen, Bertie  and
      Attanasio, Giuseppe  and
      Bianchi, Federico  and
      Hovy, Dirk},
    editor = "Duh, Kevin  and
      Gomez, Helena  and
      Bethard, Steven",
    booktitle = "Proceedings of the 2024 Conference of the North American Chapter of the Association for Computational Linguistics: Human Language Technologies (Volume 1: Long Papers)",
    month = jun,
    year = "2024",
    address = "Mexico City, Mexico",
    publisher = "Association for Computational Linguistics",
    url = "https://aclanthology.org/2024.naacl-long.301/",
    doi = "10.18653/v1/2024.naacl-long.301",
    pages = "5377--5400"
}

@misc{huang2025safetytax,
      title={Safety Tax: Safety Alignment Makes Your Large Reasoning Models Less Reasonable}, 
      author={Tiansheng Huang and Sihao Hu and Fatih Ilhan and Selim Furkan Tekin and Zachary Yahn and Yichang Xu and Ling Liu},
      year={2025},
      eprint={2503.00555},
      archivePrefix={arXiv},
      primaryClass={cs.CR},
      url={https://arxiv.org/abs/2503.00555}, 
}

@inproceedings{jiang2025safechain,
    title = "{S}afe{C}hain: Safety of Language Models with Long Chain-of-Thought Reasoning Capabilities",
    author = "Jiang, Fengqing  and
      Xu, Zhangchen  and
      Li, Yuetai  and
      Niu, Luyao  and
      Xiang, Zhen  and
      Li, Bo  and
      Lin, Bill Yuchen  and
      Poovendran, Radha",
    editor = "Che, Wanxiang  and
      Nabende, Joyce  and
      Shutova, Ekaterina  and
      Pilehvar, Mohammad Taher",
    booktitle = "Findings of the Association for Computational Linguistics: ACL 2025",
    month = jul,
    year = "2025",
    address = "Vienna, Austria",
    publisher = "Association for Computational Linguistics",
    url = "https://aclanthology.org/2025.findings-acl.1197/",
    doi = "10.18653/v1/2025.findings-acl.1197",
    pages = "23303--23320",
    ISBN = "979-8-89176-256-5",
}

@inproceedings{wang2026star1,
author = {Wang, Zijun and Tu, Haoqin and Wang, Yuhan and Wu, Juncheng and Liu, Yanqing and Mei, Jieru and Bartoldson, Brian R. and Kailkhura, Bhavya and Xie, Cihang},
title = {{STAR-1}: Safer Alignment of Reasoning {LLMs} with {1K} Data},
booktitle = {Proceedings of the AAAI Conference on Artificial Intelligence},
year = {2026},
isbn = {978-1-57735-906-7},
publisher = {AAAI Press},
url = {https://doi.org/10.1609/aaai.v40i44.41136},
doi = {10.1609/aaai.v40i44.41136},
articleno = {4235},
numpages = {10}
}

@inproceedings{
jeung2025safepath,
title={{SAFEPATH}: Preventing Harmful Reasoning in Chain-of-Thought via Early Alignment},
author={Wonje Jeung and Sangyeon Yoon and Minsuk Kahng and Albert No},
booktitle={The Thirty-ninth Annual Conference on Neural Information Processing Systems},
year={2025},
doi={10.52202/085713-3331},
url-={https://openreview.net/forum?id=vIaNnnQxcl}
}

@inproceedings{zhou2025safekey,
    title = "{S}afe{K}ey: Amplifying Aha-Moment Insights for Safety Reasoning",
    author = "Zhou, Kaiwen and Zhao, Xuandong and Liu, Gaowen and Srinivasa, Jayanth and Feng, Aosong and Song, Dawn and Wang, Xin Eric",
    editor = "Christodoulopoulos, Christos and Chakraborty, Tanmoy and Rose, Carolyn and Peng, Violet",
    booktitle = "Proceedings of the 2025 Conference on Empirical Methods in Natural Language Processing",
    month = nov,
    year = "2025",
    address = "Suzhou, China",
    publisher = "Association for Computational Linguistics",
    url = "https://aclanthology.org/2025.emnlp-main.1291/",
    pages = "25396--25412"
}

@techreport{mcdougall2025gemmascope2,
  author      = {Callum McDougall and Arthur Conmy and
                 J{\'a}nos Kram{\'a}r and Tom Lieberum and
                 Senthooran Rajamanoharan and Neel Nanda},
  title       = {{Gemma Scope 2}: Technical Paper},
  institution = {Google DeepMind},
  year        = {2025},
  month       = sep,
  url         = {https://deepmind.google/models/gemma/gemma-scope/}
}

@article{ko2026vllmhook,
  author        = {Ching-Yun Ko and Pin-Yu Chen},
  title         = {{vLLM Hook v0}: A Plug-in for Programming Model
                   Internals on {vLLM}},
  journal       = {arXiv preprint arXiv:2603.06588v1},
  year          = {2026},
  eprint        = {2603.06588},
  archivePrefix = {arXiv},
  primaryClass  = {cs.LG},
  doi           = {10.48550/arXiv.2603.06588},
  url           = {https://arxiv.org/abs/2603.06588v1}
}

@misc{chennabasappa2025llamafirewall,
      title={{LlamaFirewall}: An open source guardrail system for building secure {AI} agents}, 
      author={Sahana Chennabasappa and Cyrus Nikolaidis and Daniel Song and David Molnar and Stephanie Ding and Shengye Wan and Spencer Whitman and Lauren Deason and Nicholas Doucette and Abraham Montilla and Alekhya Gampa and Beto de Paola and Dominik Gabi and James Crnkovich and Jean-Christophe Testud and Kat He and Rashnil Chaturvedi and Wu Zhou and Joshua Saxe},
      year={2025},
      eprint={2505.03574},
      archivePrefix={arXiv},
      primaryClass={cs.CR},
      url={https://arxiv.org/abs/2505.03574}, 
}

@inproceedings{wang2026agentspec,
      title={{AgentSpec}: Customizable Runtime Enforcement for Safe and Reliable {LLM} Agents},
      author={Haoyu Wang and Christopher M. Poskitt and Jun Sun},
      booktitle={Proceedings of the 2026 IEEE/ACM 48th International Conference on Software Engineering},
      year={2026},
      url={https://arxiv.org/abs/2503.18666},
}

@INPROCEEDINGS{erlingsson2000irm,
  author={Erlingsson, {\'U}lfar and Schneider, Fred B.},
  booktitle={Proceeding 2000 IEEE Symposium on Security and Privacy. S\&P 2000},
  title={{IRM} enforcement of {Java} stack inspection}, 
  year={2000},
  volume={},
  number={},
  pages={246-255},
  doi={10.1109/SECPRI.2000.848461}}

@inproceedings{provos2003systrace,
author = {Provos, Niels},
title = {Improving Host Security with System Call Policies},
booktitle = {12th USENIX Security Symposium (USENIX Security 03)},
year = {2003},
address = {Washington, D.C.},
publisher = {USENIX Association},
month = aug,
pages = {257--271}
}

@inproceedings{garfinkel2003vmi,
  author    = {Tal Garfinkel and Mendel Rosenblum},
  title     = {A Virtual Machine Introspection Based Architecture for Intrusion Detection},
  booktitle = {Network and Distributed System Security Symposium (NDSS 2003)},
  year      = {2003},
  publisher = {The Internet Society},
  url-      = {https://www.ndss-symposium.org/ndss2003/virtual-machine-introspection-based-architecture-intrusion-detection/}
}

@inproceedings{
tramer2019slalom,
title={Slalom: Fast, Verifiable and Private Execution of Neural Networks in Trusted Hardware},
author={Florian Tram{\`e}r and Dan Boneh},
booktitle={International Conference on Learning Representations},
year={2019},
url={https://openreview.net/forum?id=rJVorjCcKQ},
}

@inproceedings{cunningham2026constitutional,
title={{Constitutional Classifiers++}: Efficient Production-Grade Defenses against Universal Jailbreaks},
author={Hoagy Cunningham and Jerry Wei and Zihan Wang and Andrew Persic and Alwin Peng and Jordan Abderrachid and Raj Agarwal and Bobby Chen and Austin Cohen and Andy Dau and Alek Dimitriev and Rob Gilson and Logan Howard and Yijin Hua and Jared Kaplan and Jan Leike and Mu Lin and Christopher Liu and Vladimir Mikulik and Rohit Mittapalli and Clare O'Hara and Jin Pan and Nikhil Saxena and Alex Silverstein and Yue Song and Xunjie Yu and Giulio Zhou and Ethan Perez and Mrinank Sharma},
year={2026},
booktitle={The Fourteenth International Conference on Learning Representations},
url={https://arxiv.org/abs/2601.04603},
}

@misc{kramar2026probes,
title={Building Production-Ready Probes For {Gemini}},
author={J{\'a}nos Kram{\'a}r and Joshua Engels and Zheng Wang and Bilal Chughtai and Rohin Shah and Neel Nanda and Arthur Conmy},
year={2026},
howpublished={arXiv preprint arXiv:2601.11516},
url={https://arxiv.org/abs/2601.11516},
}

@inproceedings{christodorescu2026agentsecurity,
title={Agent Security is a Systems Problem},
author={Mihai Christodorescu and Earlence Fernandes and Ashish Hooda and Somesh Jha and Johann Rehberger and Kamalika Chaudhuri and Xiaohan Fu and Khawaja Shams and Guy Amir and Jihye Choi and Sarthak Choudhary and Nils Palumbo and Andrey Labunets and Nishit V. Pandya},
year={2026},
booktitle={The Fortieth Annual Conference on Neural Information Processing Systems Position Paper Track},
url={https://arxiv.org/abs/2605.18991},
}

@misc{christodorescu2025sok,
title={Systems Security Foundations for Agentic Computing},
author={Mihai Christodorescu and Earlence Fernandes and Ashish Hooda and Somesh Jha and Johann Rehberger and Kamalika Chaudhuri and Xiaohan Fu and Khawaja Shams and Guy Amir and Jihye Choi and Sarthak Choudhary and Nils Palumbo and Andrey Labunets and Nishit V. Pandya},
year={2025},
howpublished={arXiv preprint arXiv:2512.01295},
url={https://arxiv.org/abs/2512.01295},
}

@misc{wu2024newera,
title={A New Era in {LLM} Security: Exploring Security Concerns in Real-World {LLM}-based Systems},
author={Fangzhou Wu and Ning Zhang and Somesh Jha and Patrick McDaniel and Chaowei Xiao},
year={2024},
howpublished={arXiv preprint arXiv:2402.18649},
url={https://arxiv.org/abs/2402.18649},
}

@inproceedings{wu2025isolategpt,
title={{IsolateGPT}: An Execution Isolation Architecture for {LLM}-Based Agentic Systems},
author={Yuhao Wu and Franziska Roesner and Tadayoshi Kohno and Ning Zhang and Umar Iqbal},
year={2025},
booktitle={Network and Distributed System Security Symposium (NDSS)},
url={https://arxiv.org/abs/2403.04960},
}

@inproceedings{debenedetti2025camel,
title={Defeating Prompt Injections by Design},
author={Edoardo Debenedetti and Ilia Shumailov and Tianqi Fan and Jamie Hayes and Nicholas Carlini and Daniel Fabian and Christoph Kern and Chongyang Shi and Andreas Terzis and Florian Tram{\`e}r},
booktitle={IEEE Conference on Secure and Trustworthy Machine Learning (SaTML)},
year={2026},
url={https://arxiv.org/abs/2503.18813},
}

@inproceedings{costa2025fides,
title={Securing {AI} Agents with Information-Flow Control},
author={Manuel Costa and Boris K{\"o}pf and Aashish Kolluri and Andrew Paverd and Mark Russinovich and Ahmed Salem and Shruti Tople and Lukas Wutschitz and Santiago Zanella-B{\'e}guelin},
booktitle={The Fortieth Annual Conference on Neural Information Processing Systems},
year={2026},
url={https://arxiv.org/abs/2505.23643},
}

@misc{shi2025progent,
title={{Progent}: Securing {AI} Agents with Privilege Control},
author={Tianneng Shi and Jingxuan He and Zhun Wang and Hongwei Li and Linyu Wu and Wenbo Guo and Dawn Song},
year={2025},
howpublished={arXiv preprint arXiv:2504.11703},
url={https://arxiv.org/abs/2504.11703},
}

@misc{beurerkellner2025designpatterns,
title={Design Patterns for Securing {LLM} Agents against Prompt Injections},
author={Luca Beurer-Kellner and Beat Buesser and Ana-Maria Cre{\c{t}}u and Edoardo Debenedetti and Daniel Dobos and Daniel Fabian and Marc Fischer and David Froelicher and Kathrin Grosse and Daniel Naeff and Ezinwanne Ozoani and Andrew Paverd and Florian Tram{\`e}r and V{\'a}clav Volhejn},
year={2025},
howpublished={arXiv preprint arXiv:2506.08837},
url={https://arxiv.org/abs/2506.08837},
}

@misc{zou2026agentkernel,
title={{AgentKernel}: The Trust-Native Agentic Operating System},
author={Zhenhua Zou and Sheng Guo and Qiuyang Zhan and Lepeng Zhao and Shuo Li and Zhuotao Liu},
year={2026},
howpublished={arXiv preprint arXiv:2609.29647},
url={https://arxiv.org/abs/2609.29647},
}

@inproceedings{wolf2020transformers,
title={Transformers: State-of-the-Art Natural Language Processing},
author={Wolf, Thomas and Debut, Lysandre and Sanh, Victor and Chaumond, Julien and Delangue, Clement and Moi, Anthony and Cistac, Pierric and Rault, Tim and Louf, Remi and Funtowicz, Morgan and Davison, Joe and Shleifer, Sam and von Platen, Patrick and Ma, Clara and Jernite, Yacine and Plu, Julien and Xu, Canwen and Le Scao, Teven and Gugger, Sylvain and Drame, Mariama and Lhoest, Quentin and Rush, Alexander},
booktitle={Proceedings of the 2020 Conference on Empirical Methods in Natural Language Processing: System Demonstrations},
pages={38--45},
year={2020},
publisher={Association for Computational Linguistics},
doi={10.18653/v1/2020.emnlp-demos.6},
url={https://aclanthology.org/2020.emnlp-demos.6/},
}

@misc{yang2025qwen3,
title={Qwen3 Technical Report},
author={Yang, An and Li, Anfeng and Yang, Baosong and others},
year={2025},
howpublished={arXiv preprint arXiv:2505.09388},
url={https://arxiv.org/abs/2505.09388},
}

@misc{gemmateam2025gemma3,
title={Gemma 3 Technical Report},
author={{Gemma Team}},
year={2025},
howpublished={arXiv preprint arXiv:2503.19786},
url={https://arxiv.org/abs/2503.19786},
}

@inproceedings{hanna2025circuittracer,
title={{Circuit-Tracer}: A New Library for Finding Feature Circuits},
author={Hanna, Michael and Piotrowski, Mateusz and Lindsey, Jack and Ameisen, Emmanuel},
booktitle={Proceedings of the 8th {BlackboxNLP} Workshop: Analyzing and Interpreting Neural Networks for {NLP}},
pages={239--249},
year={2025},
publisher={Association for Computational Linguistics},
doi={10.18653/v1/2025.blackboxnlp-1.14},
url={https://aclanthology.org/2025.blackboxnlp-1.14/},
}

@misc{alain2016probes,
title={Understanding intermediate layers using linear classifier probes},
author={Guillaume Alain and Yoshua Bengio},
year={2016},
howpublished={arXiv preprint arXiv:1610.01644},
url={https://arxiv.org/abs/1610.01644},
}

@inproceedings{fang2026nextguard,
title={{NExT-Guard}: Streaming Safeguard without Token-Level Labels},
author={Junfeng Fang and Nachuan Chen and Houcheng Jiang and Dan Zhang and Xiangnan He and Tat-Seng Chua and Xiang Wang},
booktitle={Proceedings of the 43rd International Conference on Machine Learning (ICML)},
year={2026},
url={https://proceedings.mlr.press/v306/fang26a.html},
}

@inproceedings{lee2025cast,
title={Programming Refusal with Conditional Activation Steering},
author={Bruce W. Lee and Inkit Padhi and Karthikeyan Natesan Ramamurthy and Erik Miehling and Pierre Dognin and Manish Nagireddy and Amit Dhurandhar},
booktitle={International Conference on Learning Representations (ICLR)},
year={2025},
url={https://openreview.net/forum?id=Oi47wc10sm},
}

@misc{yu2026dmilib,
title={Enabling Performant and Flexible Model-Internal Observability for {LLM} Inference},
author={Nengneng Yu and Sixian Xiong and Yibo Zhao and Wei Wang and Zaoxing Liu},
year={2026},
howpublished={arXiv preprint arXiv:2605.11093},
url={https://arxiv.org/abs/2605.11093},
}

@misc{cooney2026vllmlens,
title={{vLLM-Lens}},
author={Alan Cooney and Sid Black},
year={2026},
howpublished={\url{https://github.com/UKGovernmentBEIS/vllm-lens}},
note={{GitHub} repository, release v1.3.0, accessed 2026-11-14},
}

@misc{fang2026mia,
title={{MIA}: Runtime-Reconfigurable Access to Model Internals in {LLM} Inference Engines},
author={Yunhua Fang and Ching-Yun Ko and Sarathkrishna Swaminathan and Tong Zhang and Pin-Yu Chen},
year={2026},
howpublished={Technical report},
url={https://github.com/IBM/vLLM-Hook},
note={Manuscript in the docs folder of the repository, accessed 2026-11-14},
}

@inproceedings{souly2024strongreject,
title={A {StrongREJECT} for Empty Jailbreaks},
author={Alexandra Souly and Qingyuan Lu and Dillon Bowen and Tu Trinh and Elvis Hsieh and Sana Pandey and Pieter Abbeel and Justin Svegliato and Scott Emmons and Olivia Watkins and Sam Toyer},
booktitle={The Thirty-eighth Annual Conference on Neural Information Processing Systems Datasets and Benchmarks Track},
year={2024},
url={https://arxiv.org/abs/2402.10260},
}

@inproceedings{wu2025promptpeek,
  author    = {Guanlong Wu and Zheng Zhang and Yao Zhang and Weili Wang and Jianyu Niu and Ye Wu and Yinqian Zhang},
  title     = {I Know What You Asked: Prompt Leakage via {KV}-Cache Sharing in Multi-Tenant {LLM} Serving},
  booktitle = {Network and Distributed System Security Symposium ({NDSS})},
  year      = {2025},
  url       = {https://www.ndss-symposium.org/ndss-paper/i-know-what-you-asked-prompt-leakage-via-kv-cache-sharing-in-multi-tenant-llm-serving/}
}

@inproceedings{kuchler2026withinbatch,
title={Architectural Backdoors for Within-Batch Data Stealing and Model Inference Manipulation},
author={Nicolas K{\"u}chler and Ivan Petrov and Conrad Grobler and Ilia Shumailov},
booktitle={IEEE Conference on Secure and Trustworthy Machine Learning (SaTML)},
year={2026},
url={https://arxiv.org/abs/2505.18323},
}

@inproceedings{chen2025struq,
title={{StruQ}: Defending Against Prompt Injection with Structured Queries},
author={Sizhe Chen and Julien Piet and Chawin Sitawarin and David Wagner},
booktitle={34th USENIX Security Symposium (USENIX Security 25)},
year={2025},
url={https://arxiv.org/abs/2402.06363},
}

@inproceedings{abdelnabi2025drift,
title={Get My Drift? {C}atching {LLM} Task Drift with Activation Deltas},
author={Sahar Abdelnabi and Aideen Fay and Giovanni Cherubin and Ahmed Salem and Mario Fritz and Andrew Paverd},
booktitle={IEEE Conference on Secure and Trustworthy Machine Learning (SaTML)},
year={2025},
url={https://arxiv.org/abs/2406.00799},
}

@IEEEtranBSTCTL{IEEEexample:BSTcontrol,
  CTLdash_repeated_names = "no"
}

\appendix
This appendix describes the settings and checks behind Section~\ref{sec:evaluation}, in the same order.

\textbf{Generation and judging.}
Sampling also uses top-$p$ 0.95 and top-$k$ 20, and the three seeds are 0, 1, and 2.
Additionally, WildGuard leaves at most 1.3\% of the responses in Table~\ref{tab:filter} without a label.
The table excludes them, and each of its comparisons holds whether they count as harmful or as safe.
For the combined row, the filter reads each response that Checkpoint released, and a response that it blocks becomes the fixed refusal.
Each unchanged pair of a prompt and a response keeps its WildGuard label.
WildGuard judges the 41 new pairs of a prompt and the fixed refusal in one more pass.
All three added times come from one session on the same 128 responses, 32 from each of HarmBench, WildJailbreak, XSTest, and the ordinary requests.
In this session, the filter of the combined row reads the responses that Checkpoint releases, 52 of which are the fixed refusal.
The time that Checkpoint saves by stopping refused requests early is not subtracted.

\textbf{Mediation tests.}
Each association run limits answers to 512 tokens, which keeps requests entering and leaving the batch.
As a result, each of the 14 runs of the probe policies reuses the rows of finished requests more than 1,500 times.
These runs cover both schedules, device and host evaluation, widths of 32 and 128 requests, and bursty or staggered arrivals.
In addition, four of them enable chunked prefill and prefix caching, and eight more runs repeat the workload with the released transcoder and SAE.
To test the startup check, we start LMSM in 11 configurations outside the supported set, such as a compiled model without the capture or host evaluation with asynchronous scheduling, chunked prefill, or prefix caching.
Moreover, the failure tests bypass this check and then remove the capture of 384 requests or drop one observation of 32 requests, under both schedules and both evaluation modes.
Each of these 416 requests fails closed and releases only the fixed refusal.
To compare device and host evaluation, we run the same 64 requests through both modes for each schedule.
To check coverage, we ask for answers of one or two sentences on HarmBench, WildJailbreak, and XSTest.
With the policy set either to release such answers with a record or to refuse them, LMSM records a request as not reached exactly when its answer ends before the checkpoint.
Finally, each of the 7,668 requests that the deployed transcoder and SAE policies served was released as its decision required.

\textbf{Released-artifact rules.}
Each rule of the released transcoder and SAE fires when enough of a few named features reach their thresholds.
A search over the activations of harmful prompts from each rule's category and of benign prompts set these features and thresholds once, before our evaluation.
This search could have drawn on 100 of the 250 XSTest prompts and on the behaviors of 156 of the 602 HarmBench prompts.
Without these prompts, each category rule's firing on its own target in Figure~\ref{fig:rules} changes by at most 6 points, and its firing on XSTest by at most 2 points.
Similarly, each of the transcoder's category rules alone still reduces the ASR on its own category by half or more, while its FRR remains at or below 3.1\%.
The SAE's chemical and biological rule reduces the ASR on its category from 48.8\% to 29.3\% and increases the FRR by 1.6 points.
As a result, the prompts that the search could have drawn on do not explain where the rules fire or how much each rule removes.
The copyright requests are the HarmBench copyright behaviors outside this search, 37 for books and 37 for songs, asked plainly and answered with three seeds.
We replay the rules on these answers.
HarmBench's copyright classifier marks 1.4\% of the answers of Qwen3-4B and 4.1\% of those of Gemma-3-4B as reproducing the text.
We also ran the two rules of each artifact as a policy, and each policy's XSTest FRR is within 0.3 points of the replayed FRR.

\textbf{Rule replay.}
A rule fires when the fraction of its features whose running maximum (the largest value so far) reaches the feature's threshold is at least the rule's required fraction.
These maxima only grow, and the gate holds each response until generation ends.
As a result, the final maxima of a Disabled run decide whether each rule would have fired and what any subset of rules would have released.
This replay agrees with each of the 2,556 requests of each deployed artifact policy, and it contains the first rule that the transcoder's Disabled run recorded for each of its 1,806 requests.

\textbf{Serving cost.}
Each configuration runs one warm-up and five measured repetitions per round, and it starts only after the GPU has been idle for 20 seconds.
Moreover, a process sampler found no other job on the measured GPU during any of these repetitions.
Each line of Figure~\ref{fig:cost}c is relative to native vLLM on the same model in the same session, including the SAE's line on Gemma-3-4B.
Finally, the larger models run a Checkpoint policy with synthetic parameters, and on Qwen3-4B, the throughput of such a policy is within 0.1 points of the fitted policy's.
This agreement shows that the cost depends on the structure of the policy rather than on its parameter values.

\end{document}